\documentclass[12pt,amsfonts,enumerate,amscd]{amsart}
\usepackage{enumerate}
\usepackage{amsthm}
\numberwithin{equation}{section}

\usepackage{graphicx}

\newtheorem{thm}{Theorem}[section]

\newcommand{\theor}[1]{Theorem \ref{#1}}

\newcommand{\sect}[1]{Section \ref{#1}}
\newcommand{\subsect}[1]{Subsection \ref{#1}}

\newcommand{\bbe}{\begin{equation}}
\newcommand{\lan}{\langle}
\newcommand{\ran}{\rangle}

\newcommand{\thet}{{\theta}}
\newcommand{\tthet}{{\tilde \theta}}

\newcommand{\bP}{{\bf P}}
\newcommand{\Q}{{\bf Q}}

\newcommand{\R}{{\bf R}}
\newcommand{\C}{{\bf C}}

\newcommand{\cD}{{\mathcal D}}

\newcommand{\cC}{{\mathcal C}}

\newcommand{\cL}{{\mathcal L}}

\newcommand{\dt}{{\Delta t}}

\newcommand{\intrn}{\int_{\R^n}}
\newcommand{\rn}{\R^n}
\newcommand{\hu}{\hat u}

\newcommand{\hf}{\hat f}

\newcommand{\tf}{\tilde f}
\newcommand{\tg}{\tilde g}

\newcommand{\eps}{\epsilon}
\newcommand{\de}{\delta}
\newcommand{\al}{\alpha}
\newcommand{\be}{\beta}

\newcommand{\ee}{\end{equation}}

\newcommand{\dd}{\partial}
\newcommand{\la}{\lambda}
\newcommand{\lp}{\lambda_+}
\newcommand{\lm}{\lambda_-}

\newcommand{\om}{\omega}

\newcommand{\ka}{\kappa}
\newcommand{\sg}{\sigma}

\newcommand{\INT}{\int_{-\infty}^{+\infty}}

\newcommand{\eq}[1]{(\ref{#1})}

\begin{document}
\title[Pseudo-diffusions and QTSMS]{Pseudo-diffusions and Quadratic term structure models}

\author[S.~Levendorski\v{i}]{Sergei
Levendorski\v{i}}\thanks{The author thanks the participants of The
Summer Meeting of the Econometric Society, Evanston, June 2003,
and Mathematical Finance seminar at the University of Texas at
Austin for useful discussions. The author is especially grateful
to an anonymous referee and associated editor of the first version
of the paper, and to David Chapman, Qiang Dai, Darrell Duffie, Don
Kim and Ken Singleton for valuable comments and suggestions. The
usual disclaimer applies.\\ Address: Sergei Levendorski\v{i}, The
University of Texas at Austin, Department of Economics, 1
University Station C3100, Austin, TX, 78712-0301, e-mail
leven@eco.utexas.edu}
\maketitle



\centerline{\small Department of Economics, The University of
Texas at Austin}

\begin{abstract}
The non-gaussianity of processes observed in financial markets and
relatively good performance of gaussian models can be reconciled
by replacing the Brownian motion with  L\'evy processes whose
L\'evy densities decay as $\exp(-\la|x|)$ or faster, where $\la>0$
is large. This leads to  asymptotic pricing models. The leading
term, $P_0$, is the price in the Gaussian model with the same
instantaneous drift and variance. The first correction term
depends on the instantaneous moments of order up to three, that
is, the skewness is taken into account, the next term depends on
moments of order four (kurtosis) as well, etc. In empirical
studies, the asymptotic formula can be applied without explicit
specification of the underlying process: it suffices to assume
that the instantaneous moments of order greater than two are small
w.r.t. moments of order one and two, and use empirical data on
moments of order up to three or four. As an application, the bond
pricing problem in the non-Gaussian quadratic term structure model
is solved.

For pricing of options near expiry, a different set of asymptotic
formulas is developed; they require more detailed specification of
the process, especially of its jump part. The leading terms of
these formulas depends on the jump part of the process only, so
that they can be used in empirical studies to identify the jump
characteristics of the process.


\end{abstract}

{\sc Key words:} Quadratic term structure models, L\'evy
processes, asymptotic solutions

\newpage
\section{Introduction}

To account for
 fat tails, skewness and excessive
kurtosis of empirical probability distributions of returns in real
Financial Markets, it has become increasingly popular to model the
dynamics of market factors as a L\'evy process. L\'evy models are
more realistic than Gaussian ones but the latter are much more
tractable. Indeed, in the Gaussian framework, explicit pricing
formulas are known for a wide range of options and other
contingent claims both without and with early exercise features,
whereas in the L\'evy models, most of the pricing formulas have
been obtained for contingent claims of the European type, with the
deterministic life-span. There are some explicit analytic results
for options with early exercise features: see Boyarchenko and
Levendorski\v{i} (2000, 2001, 2002a, b), Mordecki (2002) and the
bibliography therein for pricing of perpetual American options,
and Boyarchenko and Levendorski\v{i} (2002b, c) for pricing of
barrier options and first touch digitals. However, the pricing
formulas are complicated and difficult for numerical
implementation except for a rather special case of pricing of
perpetual American options under exponential jump-diffusions or
spectrally one-sided processes.

Another obstacle for non-Gaussian modelling arises when one
considers more general Markov processes. The explicit pricing
formulas in affine term structure models and certain L\'evy-driven
Ornstein-Uhlenbeck models  are known in the case of contingent
claims with the deterministic life span only  -- see Duffie et al.
(2000, 2002), Chacko and Das (2002),  and Barndorff-Nielsen and
Shephard (2001b), Barndorff-Nielsen et al (2002), respectively;
for non-Gaussian variants of the HJM-model, see Eberlein and
Raible (1999). In the general case, the dependance on the state
variable does not allow one to obtain explicit analytical answers.

The following observation helps to obtain efficient approximate
solutions. As Barndorff-Nielsen and Levendorski\v{i} (2001)
notice,  typically, a good fit to the data can be achieved with
L\'evy processes whose L\'evy densities decay as $\exp(-\la|x|)$
or faster, where $\la>0$ (the steepness parameter of the
exponential L\'evy process) is large. They used this property to
derive an asymptotic pricing formula for European options under
certain class of Feller processes. The same observation was used
in Boyarchenko and Levendorski\v{i} (2002a,b,d) and Kudryavzev and
Levendorski\v{i} (2002) to derive efficient approximate formulas
for perpetual American and Bermudan options, and
first-touch-digitals, respectively.

It was shown in Boyarchenko and Levendorski\v{i} (2002a, b) that
the simple approximate formula is of the same form as the
corresponding formula in a Gaussian model even when the underlying
L\'evy process has no Gaussian component. It can be shown that the
leading term of the approximate pricing formula in
Barndorff-Nielsen and Levendorski\v{i} (2001) can also be written
as the pricing formula in a Gaussian model. These observations can
serve as an analytical explanation of relatively good performance
of Gaussian models in apparently non-Gaussian situations. Thus, as
far as pricing formulas are concerned, L\'evy processes with large
steepness parameters behave almost as the Brownian motion, and
Feller processes with large steepness parameters considered in
Barndorff-Nielsen and Levendorski\v{i} (2001) behave almost as
Gaussian diffusions.  It seems reasonable to use the nomer {\em
pseudo-diffusions} for L\'evy processes and more general
L\'evy-like Feller processes with large steepness parameters .

The modelling with pseudo-diffusions allows one to obtain an
efficient approximation to the price; in some situations, the
asymptotic expansion of the price can be obtained, of the form
\begin{equation}\label{asprice}
P(x, t)\sim P_0(x, t)(1+\la^{-1}P_1(x, t)+\la^{-2}P_2(x,
t)+\cdots),
\end{equation}
where the leading term, $P_0$, is the price in the Gaussian model
with the same instantaneous drift and variance. The first
correction term takes into account the moments of order three as
well (skewness), the second correction term accounts for moments
of order four, etc. Notice that though the leading term  looks as
the pricing formula in the Gaussian model, the ``drift" and
``variance-covariance matrix" used in the formula for the leading
term are not the same as the ones of the Gaussian component of the
process unless it is purely Gaussian. Indeed, a L\'evy process may
have no diffusion component at all.

The aim of the paper is  to apply the approximate pricing approach
to quadratic term structure models (QTSM) when the stochastic
factor follows a mean-reverting pseudo-diffusion process of the
simplest form (it is unlikely that in the QTSM model, an explicit
pricing formula can be obtained unless the process process is
Gaussian), and derive a pricing formula of the form \eq{asprice}.
For the discussion about advantages of the Gaussian QTSM model,
see Ahn et al (2002, 2003) and Chen and Poor (2002). Cheng and
Scaillet (2002) consider an affine-quadratic model, and  allow for
jumps but only in the dynamics of affine variables of the model.
Notice that the use of jumps in QTSM models adds additional
flexibility in joint modelling under the historic and a
risk-neutral measures, and one may hope that the performance of
QTSM models can be improved by introducing jumps.\footnote{The
author is grateful to an anonymous referee for this suggestion}
Another improvement (and quite sizable one) is expected in pricing
of out-of-the-money options near expiry, where the main
contribution to the price comes from the jump part of the process.
Near expiry, however, a different approximate formulas are needed,
which use more detailed information about the jump part of the
processes than the skewness and kurtosis. These formulas are
similar to approximate formulas for out-of-the-money options on
stocks developed in Levendorski\v{i} (2003), and can be derived by
the same reasoning.

\subsection{Plan of the paper}  In \sect{sectlevy}, we list families of
exponential L\'evy processes used in empirical and theoretical
studies of financial markets. In \sect{sectgen}, we formulate the
pricing problem for an interest rate derivative of the European
type, and by using the Feynman-Kac theorem, reduce the pricing
problem to the boundary problem for an integro-differential
equation. We also explain the scheme of the asymptotic pricing. In
\sect{sectbondG}, we recall the solution of the bond pricing
problem  in the one-factor Gaussian case, and indicate the
properties of the solution which are crucial for our asymptotic
method. In \sect{sectbond1}, we demonstrate our method in the
simplest case of the one-factor L\'evy model for the bond price,
and  present numerical examples. In \sect{exten}, we consider
possible specifications of the market price of risk, the
generalization for the multi-factor case, derive approximate
formulas for interest rate derivatives near expiry, and suggest a
procedure of parameter fitting based on the asymptotic expansions.
In \sect{concl}, we summarize our results, and compare the L\'evy
QTSM with multi-factor Gaussian QTSM. In the appendix, technical
results are proven.

\section{L\'evy processes in financial modelling}\label{sectlevy}
As early as in 1963, Mandelbrot suggested to use stable L\'evy
processes.
The modelling with stable L\'evy processes is not quite realistic
since the tails of L\'evy stable distributions are too fat
(polynomially decaying), whereas the tails of distributions of
returns observed in real financial markets exhibit exponential
decay. Moreover, the second moment of a L\'evy stable distribution
is infinite (unless it is a Gaussian one). This contradicts the
observed convergence to the Gaussian distribution over a longer
time scale, and even worse, the underlying stock itself should
have the infinite price under the stable L\'evy process, which
makes the model inconsistent for pricing purposes. Starting with
the beginning of the 90th, several families of L\'evy processes
with probability distributions having exponentially decaying tails
  have been  used
to describe the behavior of stock prices in real financial
markets:
\begin{itemize}
\item
Variance Gamma Processes (VGP) constructed and used by Madan and
co-authors in a series of papers during 90th
(see Madan et al. (1998) and the bibliography therein);

\item
Hyperbolic Processes (HP) were constructed and used
by Eberlein and co-authors (see Eberlein et al. (1998), Eberlein
and Prause (1999)); hyperbolic distributions were constructed
by Barndorff-Nielsen (1977));

\item
Normal Inverse Gaussian Processes (NIG) were introduced
by Barndorff-Nielsen (1998) and used to model German stocks
by Barndorff-Nielsen and Jiang (1998);

\item
Truncated L\'evy Processes (TLP) constructed by Koponen (1995)
  were used for modeling in real financial markets
by
Bouchaud and Potters (1997), Cont et al (1997) and  Matacz (2001);
the extended Koponen family was constructed in
Boyarchenko and Levendorski\v{i} (2000) (the generalization was
needed since probability distribution of Koponen's family have
tails of the same rate of exponential decay whereas in real
financial markets, the left tail is usually much fatter; in Carr
et al (2002) and Boyarchenko and Levendorski\v{i} (2002a,b), the
extended Koponen family is called CGMY-model and KoBoL family,
respectively).

\item
Normal Tempered Stable L\'evy processes were constructed in
Barndorff-Nielsen and Levendorski\v{i} (2001) and
Barndorff-Nielsen and Shephard (2001a); they contain NIG as a
subclass.
\end{itemize}

In Boyarchenko and Levendorski\v{i} (2000), a general class of
L\'evy processes, which contained all the classes listed above
modulo certain reservation about VGP was introduced, under the
name Generalized Truncated L\'evy Processes. Later, in
Barndorff-Nielsen and Levendorski\v{i} (2001), the name: ``Regular
L\'evy processes of exponential type" (RLPE) was suggested.  For a
more detailed exposition, see Boyarchenko and Levendorski\v{i}
(2002a, 2002b). In order to present examples, recall that a L\'evy
process can be completely specified by its characteristic
exponent, $\psi$, definable from the equality $E[e^{i\lan\xi,
X(t)\ran}]=e^{-t\psi(\xi)}$. The characteristic exponent is given
by the L\'evy-Khintchine formula
\begin{equation}\label{psi1}
\psi(\xi)=-i\lan b, \xi\ran+\frac{1}{2}\langle A\xi, \xi\rangle
+\intrn(1+i\lan\xi, y\ran {\bf 1}_{|\cdot|\le 1}(y)-e^{i\lan\xi,
y\ran})F(dy),
\end{equation}
where $A:=\Sigma\Sigma^T$ is the variance-covariance matrix  of
the Gaussian component, $b\in\rn$, and $F(dx)$ is the L\'evy
density (density of jumps), which satisfies \[ \intrn \min\{|x|^2,
1\}F(dx)<\infty.
\] Any {\em generating triplet} $A, b, F(dx)$ with these
properties defines a L\'evy process (see e.g. Sato (1999)). If
$\Sigma=0$, then we have a pure jump process.

Wide families of jump-diffusion processes are subclasses of the
class of RLPE. In the first example, we introduce the family which
is widely used in affine term structure models (see Duffie et al
(2000) and Chacko and Das (2002)).

{\it Example 2.1.} Let $X$ be a L\'evy process with the L\'evy
density
\[ F(dx)=c_+\lp e^{\lp x}{\bf 1}_{(-\infty, 0)}(x) dx +
c_-(-\lm) e^{\lm x}{\bf 1}_{(0, +\infty)}(x) dx, \]
 where $\lp>0, \lm<-1$ and
$c_\pm> 0$. Then
\[
\psi(\xi)=
\frac{\sg^2}{2}\xi^2-ib\xi+\frac{ic_+\xi}{\lp+i\xi}+\frac{ic_-\xi}{\lm+i\xi},
\]
where $\sg^2\ge 0$ and $b\in \R$ are the variance and drift of the
Gaussian component. The $\psi(\xi)$ is analytic in the strip
$\Im\xi\in (\lm, \lp)$.

{\it Example 2.2.} The characteristic exponent of a process of
KoBoL family in 1D is of the form
\begin{equation}\label{kbl1d}
\psi(\xi)=-i\mu\xi+c\Gamma(-\nu)[\lp^\nu-(\lp+i\xi)^\nu+(-\lm)^\nu-(-\lm-i\xi)^\nu],
\end{equation}
where $\nu\in (0, 2), \nu\neq 1, c>0, \lm<0<\lp$, and $\mu\in \R$;
it is analytic in a strip $\Im\xi \in (\lm, \lp)$, and
\eq{rlpe1}-\eq{rlpe2} are satisfied in this strip.

{\it Example 2.3.} The characteristic exponent of a Normal Inverse
Gaussian process in 1D is of the form
\begin{equation}\label{nig1d}
\psi(\xi)=-i\mu\xi+\de[(\al^2-(\be+i\xi)^2)^{1/2}-(\al^2-\be^2)^{1/2}],
\end{equation}
where $\nu\in (0, 2), \de>0,$ and  $\al>|\be|$; it is analytic in
the strip $\Im\xi \in (-\al+\be, \al+\be)$, and
\eq{rlpe1}-\eq{rlpe2} are satisfied in this strip, with $\nu=1$.

Since the sum of the characteristic exponents of two RLPE's is the
characteristic exponent of an RLPE, the list of model examples can
easily be expanded. For multi-dimensional examples, see
Boyarchenko and Levendorski\v{i} (2002b).

Examples 2.1--2.3 are examples of pseudo-diffusions if $\lp,
|\lm|$, and $\al\pm\be$ are large. Typically, processes observed
in empirical studies of financial markets (hyperbolic processes
and variance gamma processes including) enjoy this property.

The majority of papers on L\'evy models deal with asset pricing.
Eberlein and Raible (1999) consider the HJM-model driven by a
L\'evy process (see also Eberlein and \"Ozkan (2001)). For the
usage of jump-diffusion processes and more general L\'evy
processes in affine term structure models of interest rates, see
 Duffie et al. (2000, 2002), Chacko and Das
(2002) and the bibliography therein. Barndorff-Nielsen and
Shephard (2001b) suggested to use L\'evy-driven Ornstein-Uhlenbeck
processes for interest rate modelling purposes. For the subsequent
developments, see Barndorff-Nielsen et al (2002).

\section{The model}\label{sectgen}
\subsection{L\'evy-driven QTSM}
In the Gaussian QTSM, the instantaneous interest rate is
represented as a quadratic function of the state variables,  and
the latter are specified as diffusions.  We assume that under an
EMM chosen by the market, the SDE of the state variables can be
written as
\begin{equation}\label{sde}
dX(t)=(\tthet(t)-\ka X(t))dt + dZ(t),
\end{equation}
where $\{Z(t)\}$ is an $n$-dimensional L\'evy process,
$\tthet:\rn\to \R$ is a continuous vector-function, and $\ka$ is a
constant $n\times n$ matrix,  whose eigenvalues $\la_j$ satisfy
the condition
\begin{equation}\label{cka}
\Re \la_j>0.
\end{equation}
The interest rate is modelled as
\begin{equation}\label{req}
r(X(t))=R_0+2\lan R_1, X(t)\ran +\lan\Gamma X(t), X(t)\ran,
\end{equation}
where $R_0\in\R$, $R_1\in\rn$ are constant scalar and vector,
 $\lan\cdot,
\cdot\ran$ is the standard inner product in $\rn$, and $\Gamma$ is
a positively definite symmetric matrix. The last condition ensures
that
\[
r(X(t))=\lan\Gamma(X(t)+\Gamma^{-1}R_1),
X(t)+\Gamma^{-1}R_1\ran+R_0-||\Gamma^{-1}R_1||^2
\]
is semi-bounded from below. By choosing $R_0, R_1$ and $\Gamma$
appropriately, one can ensure any lower bound on $r(X(t))$. Notice
that if one wishes to price a derivative of a stock whose dynamics
is characterized by $X$, then one may allow $r$ to depend only on
some of the factors $X_j(t)$, say, $r=r(X_1(t),\ldots, X_m(t))$,
where $m<n$; in this case, in \eq{req}, $R_1\in \R^m$, and
$\Gamma$ is an $m\times m$ matrix.

 If $Z$ has no jump component then the bond
pricing problem reduces to a system of ODE (Riccati equations),
which can easily be solved numerically, and in the one-factor
case, even analytically. (In the multi-factor case, a system of
Riccati equations can be reduced to a linear system; for the
explicit realization in the framework of the Gaussian QTSM, see
Kim (2003)). It seems unlikely that a reasonably simple exact
solution exists for a general L\'evy process but we manage to
obtain an {\em asymptotic solution} if $X$ is a pseudo-diffusion,
that is, the L\'evy density of $Z$ decays exponentially, and the
rate of decay is large. The leading term of the asymptotics is the
price in the Gaussian model with the same instantaneous moments of
order one and two, and the correction terms are polynomials in the
factors with coefficients depending on the time to expiry. After
the leading term is found, they can be calculated recursively, by
using only integration procedures in 1D. Thus, the suggested
method is relatively simple (though in multi-factor models, the
number of additional integration procedures may be rather large;
it is important that all the integrations remain one-dimensional,
even in a multi-factor model). In the one-factor case, the first
correction term is proportional to skewness, and the second one
depends on the skewness and kurtosis; to be more precise, the
first correction is proportional to skewness, and the second one
is the sum of two terms, one of which is proportional to the
square of the skewness, and the other to the kurtosis. In many
cases, the contribution of the kurtosis is small relative to the
other terms; if we omit the last term, then the pricing formula
becomes a sum of the leading term which looks as the price in the
Gaussian model, and the correction term, which is a quadratic
polynomial w.r.t. to skewness.

 Similar formula for the forward rate and numerical examples
 show that the first correction term has a pronounced upward hump,
 if the skewness
 is negative; in the result, the corrected forward rate curve
 can be hump shaped even when the Gaussian forward rate curve
is not, and all parameters of the model are time-independent. By
changing the parameters, various shapes of the forward rate curve
can be obtained.

Empirical studies show that both skewness and kurtosis can be
fairly large, and hence, the corrections to the Gaussian price
quite sizable. Consider, for instance, the statistics for the
daily change interest rates ($dr$) from Table 1 in Das (2002).
(The table presents descriptive statistics for the Fed Funds rate
over the period January 1988 to December 1997, and the unit is 1
percent). Mean: $m=-0.0005$; standard deviation: $\sg=0.2899$;
skewness: $\la_3=0.3950$; excess kurtosis: $k_4=19.8667$. Recall
that for probability distribution $P(dx)$,
\[
m:=\lan x\ran:=\INT xP(dx),\quad \sg^2:=\lan (x-m)^2\ran,\]
\[ \la:=\lan (x-m)^3\ran/\sg^3, \quad k_4:=\lan (x-m)^4\ran/\sg^4-3,
\]
and that if $P(dx)=P_\dt(dx)$ is the probability distribution of a
L\'evy process with the characteristic exponent $\psi$,  then
\[
m(\dt)/\dt=i\psi'(0); \quad
\sg^2(\dt)/\dt=\psi^{\prime\prime}(0);\] \[
 \quad \lan
(x-m)^3\ran(\dt)/\dt=-i\psi^{(3)}(0);\quad [\lan
(x-m)^4\ran(\dt)-3\sg^4(\dt)]/\dt=-\psi^{(4)}(0).
\]
We see that the coefficients in the third and fourth terms in the
Taylor series for $\psi$ around zero
are smaller than the second one
but non-negligible whereas in the Gaussian case all coefficients
starting from the third one are zero.

The skewness and kurtosis of the process under an EMM can assume
essentially arbitrary values provided they are small w.r.t.
variance; in particular, one should expect that the skewness of
the process under EMM is negative even when the one under historic
measure is positive as in the empirical example above.  This means
that even the one-factor approximate non-Gaussian model has two
free additional parameters (albeit small) which can be used to get
a better fit to the data than in the Gaussian model. In
multi-factor models, the number of additional free parameters is
larger still.

\subsection{Reduction of a pricing problem to a boundary problem}
Consider a contingent claim with the maturity date $T$ and
payoff $g(X(T))$. Its price at time $t<T$ is given by
\begin{equation}\label{defcc}
f(X(t),
t)=E_t\left[\exp\left(-\int_t^Tr(X(s))ds\right)g(X(T))\right].
\end{equation}
(We consider the pricing under a risk-neutral measure chosen by
the market). In applications, the payoff $g$ is measurable
(usually, continuous), and it may grow at infinity. In the latter
case, additional conditions on $Z$ may be needed. For instance, if
$g$ grows not faster than an exponential:
\begin{equation}\label{boundg}
|g(x)|\le Ce^{\om|x|},
\end{equation}
where $C$ and $ \om>0$ are independent of $x$, then it suffices to assume that
there exists $\la>\om$ such that for all $\mu$ in the ball
$|\mu|< \la$, and some $t>0$,
\begin{equation}\label{cexpect1}
E[e^{\lan\mu, Z(t)\ran}]<\infty,
\end{equation}
which
implies that the tails of probability densities of the process $Z$ decay
exponentially: faster than $\exp(-\rho|x|)$, for any $\rho<\la$.

It follows from \eq{cexpect1},
that for any $\xi=\eta+i\tau\in\C^n$ from the tube domain $\rn+iU_\la:=
\{\xi\ |\ |\Im\xi|=|\tau|<\la\}$ (in the one-factor case, a tube domain
is a strip), and
any $t>0$,
\begin{equation}\label{cexpect2}
E[e^{i\lan\xi, Z(t)\ran}]<\infty.
\end{equation}
(Instead of balls $U_\la$, one can use more general open sets
containing the origin.) It is immediate from \eq{cexpect2}, that
$\psi(\xi)$ and its derivatives w.r.t. the complex argument $\xi$
are well-defined in the same tube domain $\rn+iU_\la$ (one says
that $\psi(\xi)$ is {\em analytic} in $\rn+iU_\la$), and we may
use the latter condition on $\psi$ instead of the former condition
\eq{cexpect2}. To justify the use of the Feynman-Kac formula, we
assume that $Z$ is a regular L\'evy process of exponential type
(RLPE). This means that $\psi$ admits a representation
\begin{equation}\label{rlpe1}
\psi(\xi)=-i\lan\mu, \xi\ran+\phi(\xi),
\end{equation}
where $\mu\in\rn$,  and $\phi$ satisfies the following
 condition:
there exist $c>0, \nu\in (0, 2]$ and $\nu_1<\nu$ such that as $\xi\to \infty$ in the tube domain
$\rn+iU_\la$,
\begin{equation}\label{rlpe2}
\phi(\xi)=c|\xi|^\nu + O(|\xi|^{\nu_1})
\end{equation}
(see Boyarchenko and Levendorski\v{i} (2002b)). The $\nu$ and
$U_\la$ are called the order and type of the process.

To simplify the justification of the use of the Feynman-Kac
formula, we add unnecessary condition: for any multi-index
$\al=(\al_1,\ldots, \al_n)$, there exists a constant $C_\al$ such
that for all $\xi$ in the tube domain $\rn+iU_\la$,
\begin{equation}\label{rlpe3}
|\dd^\al\phi(\xi)|\le C_\al (1+|\xi|)^{\nu-|\al|},
\end{equation}
where $|\al|=\al_1+\cdots+\al_n$. Notice that this condition holds
for all model classes of RPPE's.

In the appendix, by making use of the Feynman-Kac formula, we will
prove the following theorem.
\begin{thm}\label{thmfk1}
Let the stochastic factor satisfy \eq{sde}, \eq{cka}, \eq{req},
\eq{cexpect1}, \eq{rlpe1}, \eq{rlpe2}, and \eq{rlpe3}, let $r$ be
given by \eq{req}, and  let $g$ be a continuous function, which
admits a bound \eq{boundg}.

 Then
 a) the stochastic expression \eq{defcc} defines a continuous function $f$,
which admits an estimate
\begin{equation}\label{boundf}
|f(x, t)|\le C_1e^{\om|x|},
\end{equation}
where $C_1$ is  independent of $x$ and $t\le T$;

b) $f$ is a unique solution to the following problem:
\begin{eqnarray}\label{fk1}
(\dd_t +\lan \tthet(t)-\ka x, \dd_x\ran +L-r(x))f(x, t)&=&0,\quad t<T,\\\label{bc1}
f(x, T)&=&g(x),\end{eqnarray}
where $L$ is the infinitesimal generator of $Z$.
\end{thm}
Recall that the infinitesimal generator of the L\'evy
process $Z$, $L$, can be represented in the form
of a pseudo-differential operator (PDO) with the symbol $-\psi$: $L=-\psi(D_x)$.
A PDO $A=a(D)$ with the symbol $a$ acts on sufficiently regular
functions as follows:
\[
(Au)(x)=(2\pi)^{-n}\intrn e^{i\lan x, \xi\ran}a(\xi)\hu(\xi)d\xi,
\]
where $\hu$ is the Fourier transform of $u$:
\[
\hu(\xi)=\intrn e^{-i\lan x, \xi\ran}u(x)dx.
\]
In particular, the partial derivative $\dd_x$  is  the PDO with
the symbol $i\xi$.

\subsection{Asymptotic pricing}
The asymptotic pricing formulas will be derived under the following conditions.
Assume that the characteristic exponent of
the driving L\'evy process depends on a small parameter $\eps>0$: $\psi(\xi)=\psi(\eps, \xi)$
 and satisfies the following three conditions. First, we require that
 the $\la$ in the definition of the tube
 domain $\rn+iU_\la$ satisfies $\la>>\eps^{-1/2}$. The next two conditions
 are formulated for $\xi$ in the tube domain $\rn+iU_\la$:

 1) in the region $|\xi|>\eps^{-1/2}$, $\psi(\eps, \xi)$ admits
 an estimate
 \begin{equation}\label{lowb}
 \Re \psi(\eps, \xi)\ge c|\xi|^\nu,
 \end{equation}
 where $\nu\in (0, 2]$ and $c>0$ are independent of $(\eps, \xi)$ in the
 region;

 2) in the region $|\xi|\le \eps^{-1/2}$, $\psi(\eps, \xi)$ admits
 an asymptotic expansion: in the one-factor case,
 \begin{equation}\label{asexp1d}
\psi(\eps, \xi)=-i\mu\xi+\frac{\sg^2}{2}\xi^2-\sum_{j=3}^\infty\eps^{j-2}k_j\cdot(i\xi)^j,
\end{equation}
where the coefficients $k_j$ are uniformly bounded:
\begin{equation}\label{boundk}
|k_j|\le C\sg^2/2,
\end{equation}
where $C$ is independent of $j$;
in the multi-variate case, \eq{asexp1d} is replaced with
\begin{equation}\label{asexpnd}
\psi(\eps, \xi)=-i\lan \mu,
\xi\ran+\frac{1}{2}||\Sigma^T\xi||^2-\sum_{j=3}^\infty\eps^{j-2}k_j(i\xi),
\end{equation}
where $k_j(\xi)$ is a homogeneous polynomial of order $j$, which admits a bound
\begin{equation}\label{boundkn}
|k_j((\Sigma^T)^{-1}i\xi)|\le C|\xi|^j,
\end{equation}
where $C$ is independent of $j$.

The asymptotic solution will be found in the following sections. Here we
explain the main idea in the one-factor case.
We look for the solution in the form
\begin{equation}\label{assol}
f=f_0+\eps f_1 +\eps^2 f_2+\cdots.
\end{equation}
From \eq{asexp1d}, we can formally write
\begin{equation}\label{asoper}
L=\mu\dd_x+\frac{\sg^2}{2}\dd_x^2+\sum_{j=3}^\infty\eps^{j-2} k_j \dd_x^j,
\end{equation}
and by substituting \eq{assol} and \eq{asoper} into \eq{fk1}, we obtain a formal
equality
\begin{equation}\label{aseq}
\left(L_0+\sum_{j=1}^\infty \eps^j
L_j\right)\left(f_0+\sum_{l=1}^\infty \eps^l f_l\right)=0, \ t<T,
\end{equation}
where
\[
L_0=\dd_t+(\tthet(t)+\mu-\ka x)\dd_x+\frac{\sg^2}{2}\dd_x^2-r(x)
\]
is of the same form as the operator in the Gaussian model, and
\[
L_l=k_{l+2}\dd_x^{l+2},\quad l=1,2,\ldots.
\]
By multiplying out in \eq{aseq} and gathering terms of the same
order in $\eps$, we obtain the following
series of problems. The
leading term of the asymptotics is found from
\begin{eqnarray*}
L_0f_0(x, t)&=&0,\ t<T;\\
f_0(x, T)&=&g(x),
\end{eqnarray*}
which is the pricing problem in the Gaussian model; and the following terms
are found step by step, by solving problems
\begin{eqnarray*}
L_0 f_l(x, t)&=&-\sum_{j=1}^l k_{j+2}\dd_x^{j+2} f_{l-j}(x, t),\ \ t<T,\\
f_l(x, T)&=&0,
\end{eqnarray*}
for $l=1,2,\ldots.$ We believe that for practical purposes, it suffices to
use an approximate formula \eq{assol} with terms up to order 2; this allows one
to take into account the skewness and kurtosis. This approximate solution
can be written as
\begin{equation}\label{assol2}
f\approx f_0+\eps k_3 f_1 +(\eps k_3)^2 f_{21} +\eps^2 k_4 f_{22},
\end{equation}
where $f_1$, $f_{21}$ and $f_{22}$ solve equations
\begin{eqnarray*}
L_0 f_1(x, t)&=&-\dd_x^3 f_0(x, t),\\
L_0 f_{21}(x, t)&=&-\dd_x^3 f_1(x, t),\\
L_0 f_{22}(x, t)&=&-\dd_x^4 f_0(x, t)
\end{eqnarray*}
in the half-space $t<T$, subject to zero boundary condition. The
explicit formulas for the bond price can be found in
\sect{sectbondG} and \sect{sectbond1}. Formula \eq{assol2} may
seem somewhat inconvenient for practical applications since it
depends on the small parameter $\eps$, which is not explicitly
specified. Notice, however, that
\[
\eps k_3=-i\psi^{(3)}(0)/3!,\quad \eps^2 k_4=-\psi^{(4)}(0)/4!,
\]
and the derivatives of the characteristic exponent at 0 can be
inferred from empirical data - see Introduction. Thus, we may
write \eq{assol2} without $\eps$:
\begin{equation}\label{assol3}
f\approx f_0-i\frac{\psi^{(3)}(0)}{3!}f_1-
\left(\frac{\psi^{(3)}(0)}{3!}\right)^2
f_{21}-\frac{\psi^{(4)}(0)}{4!}f_{22}.
\end{equation}
By using \eq{assol3}, the influence of the moments of order 3 and
4 on the price can be explicitly analyzed; and this influence is
highly non-linear in $(x, t)$, since the functions in \eq{assol3}
are.

If $P:=f$ is the bond price, then we can derive from \eq{assol3}
similar approximate formulas for the yield and forward rate.

The final remark is: in order to find a current term of the
asymptotics, we have to differentiate the previous terms,
therefore an asymptotic solution with several terms  may produce
serious errors in the neighborhood of a point where the pay-off
$g$ is not sufficiently smooth. Indeed, one can hardly expect that
a formula which is polynomial in $x$, can give a high order
approximation in this case. Hence, in a neighborhood of such a
point, a different asymptotic formula should be written: see
\sect{exten}.

\section{Bond pricing: Gaussian model, one-factor case}\label{sectbondG}
In this section, we recall the solution of problem
\eq{fk1}-\eq{bc1} in the one-factor Gaussian case, when
$\psi(\xi)=-i\mu\xi+\sg^2\xi^2/2$, and
$L=\mu\dd_x+\frac{\sg^2}{2}\dd_x^2$, and indicate the properties
of the solution which are crucial for the asymptotic method to
work.

We assume that the interest rate is a quadratic function
of the stochastic factor $X(t)$:
\begin{equation}\label{req1}
r(X(t))=R_0+2R_1X(t)+X(t)^2.
\end{equation}
The dynamics of the stochastic factor, $X$, is governed by \eq{sde}, where $\tthet$ is a scalar function, and
$\ka$ is a positive scalar. The bond price is given by
\eq{defcc} with $g(x)\equiv 1$, hence it is a bounded solution to
problem \eq{fk1}-\eq{bc1} with $g(x)=1$ in the RHS of \eq{bc1}.

 Set $\tau=T-t$, $\thet(\tau)=\tthet(T-t)+\mu$,
and with some abuse of notation, write $f(x, \tau)$ instead of
$f(x, T-\tau)$. We look for the bounded solution to the problem
\begin{eqnarray}\label{fkG}
(-\dd_\tau+(\thet(\tau)-\ka)\dd_x+\frac{\sg^2}{2}\dd_x^2-r(x))f(x, \tau)&=&0,\quad \tau>0,\\\label{bcG}
f(x, 0)&=& g(x)
\end{eqnarray}
in the form
\begin{equation}\label{solG}
f(x, \tau)=\exp \Phi_0(x, \tau),
\end{equation}
where
\begin{equation}\label{defphi0}
\Phi_0(x, \tau)=A(\tau)x^2+B(\tau)x+C(\tau).
\end{equation}
By substituting \eq{solG} into \eq{fkG}, we obtain
\begin{equation}\label{fkG1}
(\exp(-\Phi_0)\cL\exp\Phi_0)(x, \tau)-r(x)=0,
\end{equation}
where
\begin{equation}\label{defcL}
\cL=-\dd_\tau+(\thet(\tau)-\ka)\dd_x+\frac{\sg^2}{2}\dd_x^2,
\end{equation}
subject to $\Phi_0(x, 0)=0$. Straightforward calculations yield the
following system of ODE with zero initial data:
\begin{eqnarray}\label{ric2}
-A'(\tau)-2\ka A(\tau)+2\sg^2A(\tau)^2-1&=&0,\\\label{ric1}
-B'(\tau)-\ka B(\tau)+ 2\sg^2A(\tau)B(\tau)+2\thet(\tau)A(\tau)-2R_1&=&0,\\\label{ric0}
-C'(\tau)+\sg^2 A(\tau)+\frac{\sg^2}{2}B(\tau)^2+\thet(\tau)B(\tau)-R_0&=&0.
\end{eqnarray}
Equation \eq{ric2} is solved by separation of variables:
\begin{equation}\label{solA}
A(\tau)=A_1A_2\frac{1-e^{\om\tau}}{A_2-A_1e^{\om\tau}},
\end{equation}
where $A_1<0<A_2$ are roots of the quadratic equation $2\sg^2
A^2-2\ka A-1=0$, and $\om=2\sg^2(A_1-A_2)<0$. $A(\tau)$ having
being found, we can calculate $B(\tau)$ from the linear equation
\eq{ric1}:
\begin{equation}\label{solB}
B(\tau)=\frac{2e^{\om_1\tau}(A_2I_1(\tau)-A_1I_2(\tau))}{(A_2-A_1)(A_2-A_1e^{\om\tau})},
\end{equation}
where $\om_1=2\sg^2A_1-\ka$, and
\begin{eqnarray*}
I_1(\tau)&=&\int_0^\tau (A_1\thet(s)-R_1)e^{-\om_1s}ds,\\
I_2(\tau)&=&\int_0^\tau (A_2\thet(s)-R_1)e^{(\om-\om_1)s}ds.
\end{eqnarray*}
If $\thet$ is independent of $\tau$, then $I_j$ can be calculated explicitly:
\begin{eqnarray*}
I_1(\tau)&=&\frac{A_1\thet-R_1}{\om_1}(1-e^{-\om_1\tau}),\\
I_2(\tau)&=&\frac{A_2\thet-R_1}{\om_1-\om}(1-e^{(\om-\om_1)\tau}),
\end{eqnarray*}
and therefore
\begin{eqnarray}\label{solBconst}
B(\tau)&=&\frac{2}{(A_2-A_1)(A_2-A_1e^{\om\tau})}\\\nonumber
&&\times \left[\frac{A_2(A_1\thet-R_1)}{\om_1}(e^{\om_1\tau}-1)
-\frac{A_1(A_2\thet-R_1)}{\om_1-\om}(e^{\om_1\tau}-e^{\om \tau})\right].
\end{eqnarray}
Finally, we find $C(\tau)$ from \eq{ric0} by integration:
\begin{equation}\label{solC}
C(\tau)=\int_0^\tau\left[\sg^2
A(s)+\frac{\sg^2}{2}B(s)^2+\thet(s)B(s)-R_0\right]ds.
\end{equation}
If $\thet$ is constant, then $C(\tau)$ can be calculated explicitly.
 In order that $B(\tau)$ and $C(\tau)$
can be calculated explicitly, $\thet$ need not to be a constant;
for instance, one can use exponential polynomials.

To end this section, we make the crucial remark on the properties of
the solution. First, from \eq{solA},
\begin{equation}\label{negA}
A(\tau)<0,\quad \forall\ \tau>0,
\end{equation}
and as $\tau\to 0$,
\begin{equation}\label{negAA}
A(\tau)\sim A_1A_2\frac{-\om}{A_2-A_1}\tau=A_1A_22\sg^2\tau=-\tau.
\end{equation}
Hence, for any $\tau\in (0, T]$, $f(x, \tau)$  decays as
$\exp(-\tau x^2)$, as $x\to \pm\infty$, and $\hf(\xi, \tau)$, the
Fourier transform of $f(x, \tau)$ w.r.t. the first argument,
decays as $\tau^{-1/2}\exp(-\xi^2/(4\tau))$, as $\xi\to\pm
\infty$. To be more specific,
\begin{eqnarray}\label{boundftr1}
\hf(\xi, \tau)&=&\INT e^{-ix\xi +
A(\tau)x^2+B(\tau)x+C(\tau)}dx\\\nonumber &=&\frac{1}{-\pi
A(\tau)}\exp[C(\tau)+(\xi+iB(\tau))^2/(4A(\tau))].
\end{eqnarray}
We conclude that for any $N$, in the region $\tau\in (0, T]$, $|\xi|>\eps^{-1/2}$,
\begin{equation}\label{boundftr2}
|\xi^N\hf(\xi, \tau)|\le C_N e^{-\xi^2/(8\tau)},
\end{equation}
where $C_N$ is independent of $\tau$ and $\xi$. Notice that the
RHS of \eq{boundftr2} is negligible.

\section{Bond pricing: L\'evy model, one-factor case}\label{sectbond1}
\subsection{The leading term of the asymptotics}
We assume that \eq{lowb}-\eq{boundk} hold. Take $\mu$ and $\sg^2$
from \eq{asexp1d}, and denote by $f_0$ the solution to the
Gaussian bond pricing problem \eq{fkG}-\eq{bcG}; it is given by
\eq{solG}, \eq{solA}, \eq{solB} and \eq{solC}, and satisfies
\eq{negA}, \eq{negAA}, \eq{boundftr1} and \eq{boundftr2}.
Introduce $f^1:=f-f_0$. Since $f_0$ and $f$ are solutions to
problems \eq{fkG}-\eq{bcG} and \eq{fk1}-\eq{bc1}, respectively,
and \eq{asexp1d} holds, we conclude that $f^1$ is the solution to
the following problem: in the half-plane $\tau>0$,
\begin{equation}\label{fk2}
(-\dd_\tau+(\thet(\tau)-\ka x)\dd_x-\psi(D_x)-r(x))f^1(x, \tau)=-\cD_1(\eps, D_x)f_0(x, \tau),
\end{equation}
where
\begin{eqnarray*}
\cD_1(\eps, \xi)&:=-&\psi(\eps, \xi)+\frac{\sg^2}{2}\xi^2-i\mu\xi\\
&=&\sum_{j=3}^\infty \eps^{j-2}k_j(i\xi)^j\\
&=&\eps\sum_{j=3}^\infty \eps^{j-3}k_j(i\xi)^j.
\end{eqnarray*}
We also have the initial condition
\begin{equation}\label{bc2}
f^1(x, 0)=0.
\end{equation}
From \eq{boundk} and \eq{boundftr2}, the following estimate for the RHS in
\eq{fk2} follows:
\begin{equation}\label{boundftr3}
|\cD_1(\eps, \xi)\hf_0(\xi, \tau)|\le C_0\eps \tau^{-1/2}\exp(-\xi^2/(8\tau)),
\end{equation}
where $C_0$ is independent of $\eps\in (0, 1)$ and $\tau\in (0, T]$. By making the
inverse Fourier transform, we obtain
\begin{equation}\label{bounddf0}
||\cD_1(\eps, D_x)f_0(x, \tau)||_{C(\R\times[0, T])}\le C\eps,
\end{equation}
where $C$ is independent of $\eps\in (0, 1)$. By
applying the Feynman-Kac theorem
to \eq{fk2}-\eq{bc2}, the representation of $f^1$
in the form of the stochastic
integral results:
\[
f^1(x, \tau)=E_{-\tau}\left[\int_{-\tau}^0
\exp\left(-\int_{-\tau}^sr(X(s'))ds'\right)\cD(\eps, D_x)f_0(x,
s)ds\right],
\]
and from \eq{bounddf0}, we derive an estimate
\begin{equation}\label{boundf1}
|f^1(x, \tau)|\le C\eps,
\end{equation}
where $C$ is independent of $\eps\in (0, 1)$, $x\in\R$ and $\tau\in (0, T]$.

\subsection{First correction term}
Estimate \eq{boundf1} shows that $f_0$ is indeed the leading term of the asymptotics
of $f$ as $\eps\to 0$, and in view of \eq{asexp1d}, it is natural to look for the
first correction term in the form $\eps f_1$, where $f_1$ is the solution to the following
problem:
\begin{equation}\label{fk3}
(-\dd_\tau+(\thet(\tau)-\ka x)\dd_x+\frac{\sg^2}{2}\dd_x^2-r(x))f_1(x, \tau)=-k_3\dd_x^3f_0(x, \tau),
\end{equation}
in the half-plane $\tau>0$, subject to
\begin{equation}\label{bc3}
f_1(x, 0)=0.
\end{equation}
We look for $f_1$ in the form
\begin{equation}\label{solf1}
f_1(x, \tau)=k_3f_0(x, \tau)\tf_1(x, \tau)=k_3e^{\Phi_0(x, \tau)}\tf_1(x, \tau),
\end{equation}
where $\Phi_0$ is given by \eq{defphi0}.  By substituting into \eq{fk3}, we obtain that
$\tf_1$ solves the problem
\begin{equation}\label{fk4}
(\cL+\sg^2(\dd_x\Phi_0)\dd_x+e^{-\Phi_0}
(\cL e^{\Phi_0})-r(x))
\tf_1
=-e^{-\Phi_0}\dd_x^3 e^{\Phi_0},
\end{equation}
in the half-plane  $\tau>0$, subject to
\begin{equation}\label{bc4}
\tf_1(x, \tau)=0,
\end{equation}
where $\cL$ is defined by \eq{defcL}. Equation \eq{fkG1} allows us to
simplify \eq{fk4}:
\[
(\cL+\sg^2(2A(\tau)x+B(\tau))\dd_x)\tf_1=-e^{-\Phi_0}\dd_x^3 e^{\Phi_0}.
\]
We calculate the operator
\begin{eqnarray*}
e^{-\Phi_0}\dd_x^3 e^{\Phi_0}&=&(e^{-\Phi_0}\dd_x e^{\Phi_0})^3\\
&=&(\dd_x+2A(\tau)x+B(\tau))^3,
\end{eqnarray*}
and by using $\dd_x 1=0$,  rewrite \eq{fk4} as
\begin{equation}\label{fk5}
\cL_1\tf_1(x, \tau)
=\tg_0(x, \tau),
\end{equation}
where
\begin{eqnarray*}
\tg_0(x, \tau)&:=&8A^3x^3+12A^2Bx^2+(12A^2+6AB^2)x+6AB+B^3,\\
\cL_1&:=&\dd_\tau+(-\thet_1(\tau)+\ka_1(\tau)x)\dd_x-\frac{\sg^2}{2}\dd_x^2,\\
\thet_1(\tau)&:=&\thet(\tau)+\sg^2B(\tau),\\
\ka_1(\tau)&:=&\ka-2\sg^2A(\tau),
\end{eqnarray*}
and $A=A(\tau)$, $B=B(\tau)$. Clearly, we may look for the
solution to \eq{fk5} in the form of a polynomial in $x$, of degree
3, with coefficients vanishing at $\tau=0$:
\begin{equation}\label{soltf1}
\tf_1(x, \tau)=a_{13}(\tau)x^3+a_{12}(\tau)x^2+a_{11}(\tau)x+a_{10}(\tau).
\end{equation}
By substituting into \eq{fk5}, we obtain a system of linear ODE:
\begin{eqnarray}\label{as13}
a_{13}'+3\ka_1a_{13} &=&8A^3,\\
\label{as12}
a_{12}'+2\ka_1a_{12} -3\thet_1a_{13}&=&12A^2B,\\
\label{as11}
a_{11}'+\ka_1a_{11} -2\thet_1a_{12}
-3\sg^2 a_{13}&=&12A^2+6AB^2,\\
\label{as10}
a_{10}'-\thet_1a_{11}
-\sg^2 a_{12}&=&6AB+B^3,
\end{eqnarray}
which can easily be integrated step by step. Namely, let
\begin{equation}\label{defka2}
\ka_2(s)=\int_0^s \ka_1(s')ds';
\end{equation}
then
\begin{eqnarray}\label{sol13}
a_{13}(\tau)&=&8e^{-3\ka_2(\tau)}\int_0^\tau e^{3\ka_2(s)}A(s)^3ds,\\
\label{sol12} a_{12}(\tau)&=&e^{-2\ka_2(\tau)}\int_0^\tau
e^{2\ka_2(s)}
(12A(s)^2B(s)+3\thet_1(s)a_{13}(s))ds,\\
\label{sol11} a_{11}(\tau)&=&e^{-\ka_2(\tau)}\int_0^\tau
e^{\ka_2(s)} (12A(s)^2+6A(s)B(s)^2\\\nonumber && \hskip4.5cm+
2\thet_1(s)a_{12}(s)+3\sg^2a_{13}(s))ds,\\
\label{sol10} a_{10}(\tau)&=&\int_0^\tau(6A(s)B(s)+B(s)^3
+\thet_1(s)a_{11}(s)+\sg^2a_{12}(s))ds.
\end{eqnarray}
Formulas \eq{solf1}, \eq{soltf1} and \eq{sol13}-\eq{sol10}
give the first order
approximation
\begin{equation}\label{solf11}
f\approx f_0\cdot(1+\eps k_3\tf_1)
\end{equation}
to the bond price. The proof similar to the one of estimate
\eq{boundf1} albeit more involved  shows that the error of
approximation \eq{solf11} is
\begin{equation}\label{boundf2}
|f(x, \tau)-f_0(x, \tau)(1+\eps k_1\tf_1(x, \tau))|\le C\eps^2 (1+|x|^2)^{3/2},
\end{equation}
where $C$ is independent of $\eps\in (0, 1)$, $x\in\R$ and $\tau\in (0, T]$.

Unlike \eq{boundf1}, we have a polynomially
growing factor $(1+|x|^2)^{3/2}$
in the RHS of \eq{boundf2}.
Notice, however, that for practical purposes, one needs to know the bond price for
small values of $r(X(t))$, hence for small values of $X(t)$, and therefore the
polynomially growing factor $(1+|x|^2)^{3/2}$ does not matter much.

\subsection{First correction term II: the derivation based on the change
of variables}
To simplify the calculation of the next terms of the asymptotics,
it is advantageous to change the variables in equations similar to \eq{fk5}:
\begin{equation}\label{changex}
x=-\thet_2(\tau)+e^{\ka_2(\tau)}y,
\end{equation}
where $\ka_2$ is given by \eq{defka2}, and $\thet_2$ is the solution to the Cauchy problem
\begin{eqnarray*}
\thet_2'(\tau)-\ka_2(\tau)\thet_2(\tau)&=&\thet_1(\tau),\\
\thet_2(0)&=&0,
\end{eqnarray*}
that is,
\[
\thet_2(\tau)=e^{\ka_2(\tau)}\int_0^\tau e^{-\ka_2(s)}\thet_1(s)ds.
\]
The same change of the variables simplifies the calculation of $\tf_1$.
Introduce  an operator $S$ by $S(f)(y, \tau)=f(x(y), \tau)$. Under the change
of variables \eq{changex},
$-\dd_\tau +(\thet_1(\tau)-\ka_1(\tau))\dd_x\mapsto -\dd_\tau$ and $\dd_x\mapsto
e^{-\ka_2(\tau)}\dd_y$, therefore
\[
\cL_2:=S^{-1}\cL_1S=\dd_\tau-\frac{\sg^2}{2}e^{-2\ka_2(\tau)}\dd_y^2,
\]
and we can rewrite \eq{fk5} in the form
\begin{equation}\label{fk55}
\cL_2 F_1(y, \tau)
=G_0(y, \tau),
\end{equation}
where $F_1=S\tf_1$, $G_0=S\tg_0$. Clearly, $G^0$ is a polynomial in $y$
of the same order as $\tg_0$:
\[
G_0(y, \tau)=G_{0, 3}(\tau)y^3+G_{0, 2}(\tau)y^2+G_{0, 1}(\tau)y+G_{0, 0}(\tau),
\]
and the coefficients $G_{0,j}$ can easily be calculated by using
formulas for the coefficients of $\tg_0$ or, better,
independently. Under the change of variables \eq{changex},
$\dd_x+2A(\tau)x+B(\tau)$ becomes
\[
\cD:=e^{-\ka_2(\tau)}(\dd_y+A_1(\tau)y+B_1(\tau)),
\]
where
\[
A_1(\tau)=2e^{2\ka_2(\tau)}A(\tau), \quad
B_1(\tau)=e^{\ka_2(\tau)}(B(\tau)-2A(\tau)\thet_2(\tau)),
\]
therefore
\begin{eqnarray}\label{defG0}
G_0&=&\cD^3\cdot 1\\
\nonumber
&=&e^{-3\ka_2}
(A_1^3y^3+3A_1^2B_1y^2+(3A_1^2+3A_1B_1^2)y+3A_1B_1+B_1^3),
\end{eqnarray}
where $\ka_2=\ka_2(\tau), A_1=A_1(\tau)$ and $B_1=B_1(\tau)$.
The solution of \eq{fk55} subject to $F_1(y, 0)=0$
is  a polynomial in $y$ of the same order as $G_0$:
\[
F_1(y, \tau)=F_{1,3}(\tau)y^3+F_{1,2}(\tau)y^2+F_{1,1}(\tau)y+F_{1,0}(\tau),
\]
whose coefficients can easily be found by integration:
\begin{eqnarray}\label{solF13}
F_{1,3}(\tau)&=&\int_0^\tau G_{0, 3}(s)ds,\\\label{solF12}
F_{1,2}(\tau)&=&\int_0^\tau G_{0, 2}(s)ds,\\\label{solF11}
F_{1,1}(\tau)&=&\int_0^\tau (G_{0, 1}(s)+3\sg^2e^{2\ka_2(\tau)}G_{0,3}(s))ds,\\\label{solF10}
F_{1,0}(\tau)&=&\int_0^\tau (G_{0, 0}(s)+\sg^2e^{2\ka_2(\tau)}G_{0,2}(s))ds.
\end{eqnarray}
After that we make the inverse change of variables $y=e^{-\ka_2(\tau)}(x+\thet_2(\tau))$,
and calculate $\tf_1=S^{-1}F_1$.

\subsection{Next terms of the asymptotics}
Suppose that the approximation of order $j\ge 1$ has been found:
\[
f=f_0\cdot\left(1+\sum_{l=1}^j\eps^l
k_{l+2}\tf_l\right)=f_0\sum_{l=0}^j\eps^l k_{l+2}\tf_l,
\]
where $k_2=1, f_0=e^{\Phi_0}, \tf_0\equiv 1$, and $\tf_l$, $1\le
l\le j$, are polynomials in $x$ with coefficients depending on on
$\tau$:
\begin{equation}\label{deftfl}
\tf_l(x, \tau)=\sum_{s=0}^{m_l} a_{ls}(\tau)x^s.
\end{equation}
We look for the next term of the asymptotics in the form $\eps^{j+1}k_{j+3}f_0\tf_{j+1}$,
where $f_{j+1}:=f_0\tf_{j+1}$ is the solution to the problem
\begin{eqnarray}\label{fk6}
(\cL-r(x))f_{j+1}(x, \tau)&=&-g_{j}(x, \tau),\ \tau>0,\\
\label{bc6}
f_{j+1}(x, 0)&=&0,
\end{eqnarray}
where $\eps^{j+1}g_{j}$ is the collection of terms of order $\eps^{j+1}$ in the expression
\[
\sum_{p=3}^\infty \eps^{p-2}k_p\dd_x^p \left(\sum_{l=0}^j\eps^l
k_{l+2}f_0\tf_l\right),
\]
that is,
\[
g_{j}=\sum_{p=3}^{j+3} k_p k_{j+5-p}\dd_x^p(f_0\tf_{j+3-p}).
\]
Set
\begin{eqnarray}\label{deftg1}
\tg_{j}&=&e^{-\Phi_0}g_{j}\\
\nonumber
&=&\sum_{p=3}^{j+3}k_p k_{j+5-p}(e^{-\Phi_0}\dd_x e^{\Phi_0})^p\tf_{j+3-p}\\
\nonumber
&=&\sum_{p=3}^{j+3}k_p k_{j+5-p}(\dd_x+2A(\tau)x+B(\tau))^p\tf_{j+3-p},
\end{eqnarray}
and multiply \eq{fk6} by $e^{-\Phi_0}$. We obtain
\begin{equation}\label{fk66}
(\cL+\sg^2(\dd_x\Phi_0)\dd_x+e^{-\Phi_0}(\cL e^{\Phi_0})-r(x))\tf_{j+1}=
-\tg_{j}.
\end{equation}
Equation \eq{fkG1} allows us to simplify \eq{fk66} and obtain a problem in the half-space
$\tau>0$ with the unknown $\tf_{j+1}$:
\begin{eqnarray}\label{fk7}
\left(-\dd_\tau+(\thet_1(\tau)-\ka_1(\tau)x)\dd_x+\frac{\sg^2}{2}\dd_x^2\right)\tf_{j+1}&=&-\tg_{j},
\ \tau>0,\\
\label{bc7}
\tf_{j+1}(x, 0)&=&0.
\end{eqnarray}
From \eq{deftg1}, $\tg_{j}$ is a polynomial in $x$ as well:
\begin{equation}\label{deftg2}
\tg_{j}(x, \tau)=\sum_{s=0}^{m'_{j}}b_{j, s}(\tau)x^s,
\end{equation}
where $m'_{j}=\max_{0\le l\le j}(m_l-l+j+3)$, and any of the coefficients $b_{j, s}$
may be zero, that is, the order of $\tg_{j}$ may be less than $m'_{j}$.
Denote by $m_{j}$ the order of $\tg_{j}$.

By making the change of variables \eq{changex}, we simplify \eq{fk7}:
\begin{equation}\label{fk77}
\cL_2 F_{j+1}=G_j,
\end{equation}
where $F_j=S\tf_j, G_j=S\tg_j$.  Since $G_j$ is a polynomial in $y$ of order $m_j$:
\[
G_j(y, \tau)=\sum_{l=0}^{m_j}G_{j, l}(\tau)y^l
\]
(the coefficients $G_{j, l}$ will be calculated in \subsect{subcalcG}), $F_{j+1}$ also is:
\[
F_{j+1}(y, \tau)=\sum_{l=0}^{m_j}F_{j+1, l}(\tau)y^l,
\]
and the coefficients $F_{j+1, l}$ are easily found by integration:
\begin{eqnarray}\label{solF1}
F_{j+1, l}(\tau)&=&\int_0^\tau G_{j, l}(s)ds,\ l=m_j, m_j-1;\\
\label{solF2} F_{j+1, l}(\tau)&=&\int_0^\tau
\left(\frac{\sg^2}{2}e^{2\ka_2(s)} (l+2)(l+1)F_{j+1, l+2}(s)
+G_{j, l}(s)\right)ds,
\end{eqnarray}
for $l=m_j-2, m_j-3,\ldots 0.$
\subsection{Calculation of $G_{j, l}$}\label{subcalcG}
 We can rewrite \eq{deftg1} as
\begin{equation}\label{defG}
G_{j}=\sum_{p=3}^{j+3}k_p k_{j+5-p}\cD^p F_{j+3-p}, \quad j=0, 1,\ldots.
\end{equation}
By using the initial data $F_0=1$, we find $G_j$ and $F_{j+1}$ from \eq{defG}
and \eq{solF1}-\eq{solF2} step by step. In particular, $G_0$ is given by \eq{defG0},
and
\begin{equation}\label{defG1}
G_1=k_3^2G_{11}+ k_4G_{12},
\end{equation}
where
\begin{equation}\label{defG112}
G_{11}=\cD^3 F_1,\quad {\rm and}\  G_{12}=\cD^4 F_0=\cD^4\cdot 1
\end{equation}
are polynomials of degree 6 and 4, respectively.
Hence,
\begin{equation}\label{defF2}
F_1=k_3^2 F_{21}+k_4 F_{22},
\end{equation}
where $F_{21}$ and $F_{22}$ are polynomials of degree 6 and 4,
respectively, which solve \eq{fk77} with $G_{11}$ and $G_{12}$ in
the RHS (and satisfy the initial condition $F_{2j}(y, 0)=0$). The
reader can use \eq{solF1}-\eq{solF2} to obtain explicit formulas
for coefficients of $F_{2j}$ in terms of coefficients of $G_{1j}$.
Notice that though the latter can be written explicitly, in
practical implementation of the method, it is simpler to write a
program which calculates the coefficients of a polynomial $\cD P$,
given coefficients of a polynomial $P$, and use this program to
calculate $G_{1j}$ (and $G_l$ for $l>1$ should one wish it, though
in applications, it seems not a reasonable thing to do).

\subsection{Second order approximation for the bond price,
yield and forward rate} We see that the second order approximation
can be written in the form
\begin{equation}\label{second1}
f\approx e^{\Phi_0}(1+\eps k_3 \tf_1 +(\eps k_3)^2\tf_{21}+\eps^2k_4 \tf_{22}),
\end{equation}
where $\Phi_0$, $\tf_1=S^{-1}F_1$, $\tf_{21}=S^{-1}F_{21}$ and
$\tf_{22}=S^{-1}F_{22}$ are polynomials in $x$ with coefficients
depending on $\tau$ - and parameters $\tthet$, $\ka$,
$\mu:=i\psi'(0)$, $\sg^2:=\psi''(0)$. For practical
applications,\eq{second1} can be rewritten in the form
\begin{equation}\label{second2}
f\approx e^{\Phi_0}\left(1+K_3\tf_1 +
K_3^2\tf_{21}+K_4\tf_{22}\right),
\end{equation}
where
\begin{eqnarray*}
K_3:&=&\eps k_3=-i\psi^{(3)}(0)/3!,\\
K_4:&=& \eps^2 k_4=-\psi^{(4)}(0)/4!
\end{eqnarray*}
can be inferred from the data.

 Denote by $P:=f$ is the price of the bond, and by
 $P_0:=\exp(\Phi_0)$
 the price in the Gaussian model; then \eq{second2} becomes
 \begin{equation}\label{second3}
 P\approx P_0 + K_3 P_1 + K_3^2 P_{21}+K_4 P_{22},
 \end{equation}
 where
 \[
 P_1=P_0\tf_1,\ P_{21}=P_0\tf_{21}, \ P_{22}=P_0\tf_{22}.
 \]
 By using the formulas for the yield
 \[
 R(x, \tau)=-\frac{\ln P(x, \tau)}{\tau}
 \]
 and forward rate
 \[
 F(x, \tau)=-\frac{\dd }{\dd\tau}\ln P(x,\tau),
 \]
 we obtain approximate formulas
 \begin{equation}\label{yield}
 R\approx R_0 + K_3 R_1 + K_3^2 R_{21}+K_4 R_{22},
 \end{equation}
 where
 \begin{eqnarray*}
 R_0(x, \tau)&=&-\Phi_0(x, \tau)/\tau,\\
 R_1(x, \tau)&=&-\tf_1(x, \tau)/\tau,\\
 R_{21}(x,\tau)&=& (\tf_1(x, \tau)^2/2-\tf_{21}(x,\tau))/\tau,\\
 R_{22}(x, \tau)&=&-\tf_{22}(x, \tau)/\tau,
 \end{eqnarray*}
 and
 \begin{equation}\label{frate}
 F\approx F_0 + K_3 F_1 + K_3^2 F_{21}+K_4 F_{22},
 \end{equation}
 where
 \begin{eqnarray*}
 F_0(x, \tau)&=&-\frac{\dd }{\dd\tau}\Phi_0(x, \tau),\\
 F_1(x, \tau)&=&-\frac{\dd }{\dd\tau}\tf_1(x, \tau),\\
 F_{21}(x,\tau)&=& \frac{\dd }{\dd\tau}
 (\tf_1(x, \tau)^2/2-\tf_{21}(x,\tau)),\\
 F_{22}(x, \tau)&=&-\frac{\dd }{\dd\tau}\tf_{22}(x, \tau).
 \end{eqnarray*}

 \subsection{Numerical examples}\label{numerics} (The author thanks Nina Boyarchenko
for writing the programs for numerical examples and checking the
algebra in the previous two sections.) We take the simplest model
for $r$: $r=x^2$, and constant $\tthet(t)=0.06$, $\ka=0.3$,
$\mu=0$ and $\sg^2=0.08$. We also fix $x=0.25$, and study shapes
of correction terms to the bond price, yield and forward rates in
\eq{second3}, \eq{yield} and \eq{frate} (see Figures 1--3).

 From Figure 3, we clearly see that it is the first
 correction term, $F_1$, that can account for
 the hump of the forward rate curve - provided the skewness is negative
 and not too small in modulus. In the next three
 Figures 4--6, we plot the bond price, yield and forward rate;
 first, the leading term (dots), then the formula
 with the first correction term taken into account
 (solid line), and finally, the formula with the two correction
 terms (dotted line). We take the same parameters as above, and
 $K_3=-\sg^2/16=-0.005$,
 $K_4=K_3/20=2.5\cdot 10^{-4}$.
We see that fairly large skewness does produce a hump-shaped
forward rate curve, when the Gaussian curve has no hump; the
asymptotic formulas are applicable since $K_3$ is small w.r.t.
$\sg^2$, and $K_4$ is small w.r.t. $K_3$.

In the last series of figures (Figures 7--9), we fix small  $K_4=
\sg^2/200$, and show how the shapes of the curves vary with the
skewness.

Notice that if one fits the Gaussian QTSM and the non-Gaussian one
to the same data set, then the leading Gaussian-like leading term
in the L\'evy model will be determined by slightly different drift
and volatility parameters and the spot value of the factor, hence
the pictures above do not describe quite accurately the difference
between the Gaussian and L\'evy models. However, the shapes of the
curves in the Gaussian model and the leading Gaussian
approximation in the non-Gaussian models are the same (and not
differ much). Thus, one can use the pictures to get a feeling what
a difference between the Gaussian and non-Gaussian QTSMs can be.

\section{Extensions and ramifications}\label{exten}
\subsection{Multi-factor case} All the constructions in the
previous two sections admit modifications for the multi-factor
case. The only two differences are
\begin{itemize}
\item
in the Gaussian approximation, the $A(\tau)$, $B(\tau)$ and
$C(\tau)$ are matrix functions. The first two can be found by
solving a system of linear ODE's (for detailed realization in the
Gaussian QTSM-model, see Kim (2003)), and then $C(\tau)$ is found
by the integration;
\item
the correction terms are polynomials not in one factor but in
several factors. This leads to systems of linear ODE's whose
unknowns are coefficients at factors $x^\al=x^{\al_1}_1\cdots
x^{\al_n}_n$ of order $\al=\al_1+\cdots+\al_n\le m_j$, where $m_j$
depends on the step of the method.
\end{itemize}
Certainly, the systems of linear ODE to be solved become
significantly larger but they remain linear systems; therefore,
the numerical solution is fairly stable.

\subsection{Pricing under historic and risk-neutral measures}
Assume that under the historic measure, the dynamics of $X_t$ is
given by
\begin{equation}\label{sdep}
dX(t)=(\tthet(t)-\ka X(t))dt + dZ^\bP(t),
\end{equation}
where $\{Z^\bP(t)\}$ is an $n$-dimensional L\'evy process with the
characteristic exponent $\psi^\bP$. We assume that $\psi^\bP$
admits the analytic continuation into the tube domain
$\rn+iU^\bP$, where $U^\bP$ is an open set containing 0. To
specify the interest rate dynamics under a risk-neutral measure,
$\Q$, we consider first the state price deflator in the form
$\pi_t=\exp(q_t)$, where $q_t$ obeys the SDE
\begin{equation}\label{sdeq}
dq_t=-r_t dt -\langle \Lambda, dZ^\bP(t)\rangle-\Lambda_0t.
\end{equation}
The vector $\Lambda\in\rn$ represent the market prices of risk of
the factors. Notice that for processes with jumps, one cannot use
an arbitrary $\Lambda$. For the bond to be priced, it is necessary
that $\Lambda\in \overline{U^\bP}$, and the condition
$\Lambda\subset U^\bP$ suffices for the bond and
 options on the yield to be priced (to be more specific, any
 payoff which admits a polynomial bound with respect to factors is
 admissible; if a payoff grows exponentially, then additional restrictions on $\Lambda$ must be imposed).

 Set $\psi^\Q(\xi)=\psi^\bP(\xi+i\Lambda)-\psi^\bP(i\Lambda)$. It
 is easy to check that under the condition $\Lambda\in U^\bP$,
 $\psi^\Q$ is the characteristic exponent of a L\'evy process,
 call it $Z^\Q$. The new measure $\Q$ is the Esscher transform of $\bP$ popular in the literature
 on the pricing of options on stocks (see e.g. Eberlein et al. (1998) and
 Boyarchenko and Levendorski\v{i} (2002b)). Notice that $\psi^\Q$
 is analytic in the tube domain
 $U^\Q=U^\bP-\Lambda\supset\{0\}$, and choose
 $\Lambda_0=-\psi^\bP(i\Lambda)$. The straightforward calculations
 show that the pricing formula
 \begin{equation}\label{prp}
 f(x, t)=E_t^\bP\left[\frac{\pi_t}{\pi_T}g(X(T))\ |\ X(t)=x\right]
 \end{equation}
 can be written as the pricing formula under the risk-neutral
 measure $\Q$:
 \begin{equation}\label{prq}
 f(x, t)=E_t^\Q\left[\exp\left(-\int_t^T r_s ds\right)g(X(T))\ |\
 X(t)=x\right].
 \end{equation}
 It is possible that in some situations, the specification of risk \eq{sdeq}
 is not
 sufficiently flexible, and in the Gaussian QTSM, Ahn et al.
 (2002) consider a more general model for prices of risk. This more general
 specification is not applicable in L\'evy models with exponentially decaying
 L\'evy densities; however, there is an additional flexibility in
 model with jumps, which may provide any number of additional degrees
 of freedom: one can use the pricing formula \eq{prq} with
 any L\'evy process $Z^\Q$ provided the difference of L\'evy
 densities of $\bP$ and $\Q$  is of finite variation (this can be shown as in the
 case of pricing of derivatives on a stock; see e.g. Carr et al.
 (2002)). An additional restriction should be taken into account if we want to use
 the approximate formulas of Sections 4--5:  the L\'evy densities
 under $\bP$ and $\Q$
 should decay exponentially, and sufficiently fast.

 Notice that contrary to the change of measure in the Gaussian model,
 the change of measure in the non-Gaussian model may lead to the
 changes (albeit small) in the instantaneous moments  of order
 two.

\subsection{Option pricing} Let $f(x, t)$ be the price of an
European style derivative contract with the pay-off $g(X(T_1))$,
where $T_1< T$. A typical example is a call option on the yield
with the pay-off $g(X(T_1))=\max\{R(X(T_1)-K, 0\}$. For a fixed
$T_1<T$, the formulas in the preceding section allows one to find
an approximation to $g(X(T_1))$ as a function of the spot values
of factors $x=X(t)$ (and of $\tau=T-T_1$). In the Gaussian model,
this is a quadratic function, and in the non-Gaussian model, it is
a fourth-order polynomial with small coefficients of order 3 and
4. Hence, we can calculate the two roots of the equation $R(x)=K$,
which are close to the two roots in the Gaussian approximation, by
using simple perturbation technique. Denote these roots by
$x^\pm_\eps(K)$, and represent the pay-off function $g(x)$ in the
form
\[
g(x)=g^+(x)+g^-(x),
\]
where
\[
g^+(x)={\bf 1}_{[x^+_\eps, +\infty)}(x)(R(x)-K),
\]
\[
g^-(x)={\bf 1}_{(-\infty, x^-_\eps]}(x)(R(x)-K).
\]
By using the perturbation technique once again, we can calculate
the Fourier transforms of functions $g^\pm$, and reduce the
problem of the calculation of $f(x, t)$ to the family of problems
considered in Sections 4--5, with the pay-off $e^{ix\xi}$ instead
of 1. The modification of the asymptotic calculations is
straightforward albeit lengthy. In the end, we make the inverse
Fourier transform, and obtain an asymptotic formula for the price
$f(x, t)$, $t<T_1$. For a numerical realization of this formula,
one needs to choose an appropriate grid in the frequency domain,
and for each $\xi$ from the grid, solve a number of systems of
ODE's. The inverse FFT finishes the job. Notice that this is a
variant of the standard transform method (see e.g. Duffie et al.
(2000), Chacko and Das (2002) and the bibliography therein).

When using this scheme, one should remember that due to the
non-smoothness of the pay-off at the money, the approximate
formulas will not work well near expiry, especially near expiry
and strike. Fortunately, near the strike, a different
approximation -- and much simpler one -- can be derived. Fix
$x_0$, and introduce
\begin{equation}\label{rx0}
r(x_0; x)=R_0+R'(x_0)(x-x_0).
\end{equation}
 Denote by $f^0(x_0; x, t)$ the solution to the affine model
 with the short rate modelled by \eq{rx0}, and the same dynamics of the factors
 under the risk-neutral measure and pay-off as in the initial QTSM.
\begin{thm}\label{nearexpfirst}
$f(x_0, t)=f^0(x_0; x_0, t)+o(\tau)+O(\eps^\infty)$ as
$\tau=T_1-t\to+0 $ and $\eps\to 0$.
\end{thm} The proof of \theor{nearexpfirst} is relatively
straightforward, and it relies on the exponential decay of the
density of jumps. Since the solution in the affine model with
jumps is well-known and fairly simple, we get an efficient
approximation to the option price in the non-Gaussian QTSM. Should
one wish it, the correction term to the formula $f(x_0, t)\sim
f^0(x_0; x_0, t)$ can be obtained as the price of a derivative
security (in the same affine term structure model), which pays the
stream of dividends at rate $g(x)$. For in-the-money options, one
can use a constant function
 $g(x_0)$ instead of $g(x)$.

Finally, for out-of-the-money options, a simpler approximation can
be derived, in the form
\begin{equation}\label{asympnearexp}
f(x, t)\sim \cC(g; x)\tau,
\end{equation}
where $\cC(g; x)$ depends only on the L\'evy density $F(dx)$ of
the process but not on the Gaussian part and drift. For the case
of the call option on the yield considered above, essentially the
same calculations as in Levendorski\v{i} (2003) give for $x\in
(x^-_\eps, x^+_\eps)$
\begin{equation}\label{asympC}
\cC(x, t)=\INT g(x+y)F(dy),
\end{equation}
and the explicit analytic expression in terms of parameters of
model classes of RLPE processes can be derived (the modification
of the calculations in Levendorski\v{i} (2003) made for
$g(x)=(e^x-1)^+$ and $g(x)=(1-e^x)^+$ is straightforward).

\subsection{Parameters' fitting}
 Far from expiry, the leading term of the yields depends on the
instantaneous moments of order one and two, and for
out-of-the-money options on yields, near expiry, the leading term
depends only on the jump part of the process. This allows us to
suggest the following scheme of the fitting the model to the data,
under a risk-neutral measure.

\begin{enumerate}[1.]
\item
 Infer parameters of the Gaussian model (including the spot values of the factors)
 from the data on yields. Here one can use the efficient method of
 moments
 as in Ahn et al. (2002) or an estimation method integrated with the extraction of
 the state variables
 (the extended-Kalman-filter-based quasi-maximum-likelihood estimate) as in  Kim (2003), say.
  We regard these parameters' values as a zero-order
 approximation to the spot values of factors, drift, mean reverting and
 variance-covariance parameters of the non-Gaussian model.
 \item
 Given the spot values of the factors, one can infer the
 conditional characteristic function of the process from empirical
 data as in
 Singleton (1999) and calculate the moments up to order 4, or infer these moments as in
  Collin-Dufresne et al.
(2003).  However, in order to identify the contribution of jumps
more accurately, the following steps seems to be reasonable.
\item
Choose a parametrized model for the L\'evy density (or one of the
standard L\'evy models described in Section 2), and by using the
prices on interest rate derivatives near expiry and approximate
formulas near expiry, fit the parameters of the L\'evy density,
and calculate the instantaneous moments of order 3 and 4.
\item
 Calculate the correction terms to the
yields in the Gaussian approximation by using the asymptotic
formulas of Section 5 and the zero-order approximations for the
parameters in the Gaussian approximation, and moments of order 3
and 4.
\item
Subtract the correction terms from the data, and use the new data
set in the Gaussian procedure to infer the corrected values of the
spot factors and the first two moments.
\item
 Step 3 can be repeated in order to get a corrected specification of the L\'evy density and moments of
 order 3 and 4.
 \end{enumerate}
 Notice that in order that this procedure be consistent,
the resulting moments of order 3 and 4 must be small relative to
 the smallest eigenvalue of the matrix of the instantaneous second moments.

 \section{Conclusion}\label{concl}
 We constructed a class of QTSM models with
 a regular L\'evy process of exponential type
 in place of the Gaussian one in standard
 QTSM. By using the Feynman-Kac formula, we have reduced
 the pricing problem for an interest rate derivative
 to a boundary problem for a pseudo-differential operator.
 In the case of the bond, we have found an approximate solution to the boundary problem
 assuming that the tails of probability densities of a
 process decay sufficiently fast. The leading term of the approximate
 solution looks as in the Gaussian
 model (even when the underlying process has no Gaussian
 component), and the correction terms depend on skewness  and kurtosis.

 Numerical examples are produced to show that
 by changing skewness and kurtosis, various shapes of the forward rate
 curve can be obtained. In particular, negative skewness
 can produce a hump-shaped forward rate curve even when
the Gaussian curve has no hump: the very non-Gaussianity of the
process is (one of) causes of the hump of the forward rate curve.
Bond prices and the yield curve also change but the types of the
shape of the curves remain essentially the same.

We discussed possible choices of a risk-neutral measure,  and
indicated additional flexibility which the usage of jumps
provides. A brief outline of asymptotic pricing in the
multi-factor case, and the pricing of interest rate derivatives is
given. We derived simple asymptotic formulas for option prices
near expiry, and suggested a procedure of the fitting of the model
to the data.

Notice that the use of pseudo-diffusions for option pricing
instead of Gaussian models with approximately the same number of
parameters is advantageous not only near expiry where the latter
are expected to produce serious errors. If the number of
parameters is approximately the same, then the number of factors
in the former is smaller than in the latter, therefore the use of
the inverse FFT simplifies significantly in the non-Gaussian
model, and the numerical procedure becomes much more stable.

\appendix
\section{Proofs of technical results}

\subsection{Proof of \theor{thmfk1}}
By using the decomposition $g=g_+-g_-$, where $g_+(x):=\max\{g(x), 0\}$ and $g_-=g_+-g$ are non-negative,
we see that it suffices to prove \theor{thmfk1} for  continuous non-negative
$g$. Fix $\chi\in C^\infty(\R)$ such that $0\le \chi(x)\le 1$ for all $x$,
$\chi(x)=1, x\le 1$, $\chi(x)=0, x\ge 2$, and for any $m>0$, set
$g^m(x):=\chi(|x|/m)g(x)$. Then $g^m$ is a continuous function with the compact
support. Define $f^m$
by \eq{defcc} with $g^m$ in the RHS. For any $x$,
$g^m(x)\uparrow g(x)$ as $m\to \infty$, and by the Monotone Convergence Theorem,
$f^m(x)\uparrow f(x)$. Notice that $f^m\to f$ in the sense of generalized
functions: for any non-negative $u\in C^\infty_0(\rn\times (0, T))$,
\[
\intrn f^m(x)u(x)dx\to \intrn f(x)u(x)dx,\quad m\to \infty.
\]
Below we will show that
\begin{enumerate}[(i)]
\item
let $g$ be continuous and satisfy \eq{boundg}; then problem \eq{fk1}-\eq{bc1}
has a unique continuous solution $f(g; x, t)$, which satisfies \eq{boundf};
\item
in the half-space $t<T$, $f$ is of the class $C^{2, 1}$ w.r.t.
$(x, t)$;
\item
$f(g^m; \cdot, \cdot)\to f(g; \cdot, \cdot)$ as $m\to \infty$, in the sense of
generalized functions;
\item
if $g$ is a continuous function of the compact support,
then $f(x, t):=f(g; x, t)$ is given by \eq{defcc}.
\end{enumerate}
By (iv), $f(g^m; \cdot, \cdot)=f^m(\cdot, \cdot)$, by (iii) and (ii), the limit in (iii)
is a continuous function, and since we already know that $f^m(x, t)\to f(x, t)$
pointwise, we conclude that $f(\cdot, \cdot)=f(g; \cdot, \cdot)$ solves the problem \eq{fk1}-\eq{bc1},
and finish the proof of \theor{thmfk1}.

It remains to prove (i)-(iv). We start with the proof of (i).
Assume that $\ka$ is diagonalizable: there exists a matrix $C$
such that $\ka_C:=C^{-1}\ka C$ is a diagonal matrix with the
diagonal entries $\ka_j$ (if $\ka$ is not diagonalizable, an
additional step is to be made - see the end of the proof of (i)).
By making the change of variables $x=Cy$, we reduce to the case of
the diagonal matrix $\ka(=\ka_C)$; the $\psi(\xi)$ becomes
$\psi_C(\xi):=\psi((C')^{-1}\xi)$. To simplify the notation below
(and without loss of generality), we assume that $\ka$ itself is
diagonal. In \eq{fk1}-\eq{bc1}, change the variables:
\[
t=T-\tau, \quad x_j=-\tthet_{2j}(\tau)+e^{\ka_j \tau}y_j,\ j=1,\ldots, n,
\]
where
\[
\tthet_{2j}(\tau)=e^{\tau\ka_j}\int_0^\tau e^{-s\ka_j}\tthet_j(T-s)ds
\]
is the solution to ODE
\[
\tthet'_{2j}(\tau)-\ka\tthet_{2j}(\tau)=\tthet_j(T-\tau),
\]
subject to $\tthet_{2j}(0)=0$,
and set
\begin{eqnarray*}
v(y, \tau)&=&f(x, t), \\
e^{\tau\ka }&=&{\rm diag}(e^{\ka_j \tau}),\\
r_{1}(y, \tau)&=&r(-\tthet_2(\tau)+e^{\tau\ka}y).
\end{eqnarray*}
 We obtain
\begin{eqnarray}\label{afk3}
(\dd_\tau + \psi(e^{-\ka\tau}D_y)
+r_{1}(y, \tau))v(y, \tau)&=&0,\ \tau>0,\\
\label{abc3}
v(y, 0)&=&g(y).
\end{eqnarray}
Notice that $\tthet_1\in C^2([0, T])$ since $\tthet\in C([0, T])$,
and $r_1$ satisfies estimate
\begin{eqnarray}\label{estr2der}
|\dd^{\al}_y\dd^{s}_\tau r_1(y, \tau)|&\le& C_{\al, s}(1+|y|)^{2-|\al|},\quad |\al|, s=0,1,2,\\
\label{estr2}
c_0|y|^2-C_0&\le &r_1(y, \tau),
\end{eqnarray}
where $c_0>0$ and $C_0, C_{\al, s}$ depend on $T$ but not on $x\in\rn$ and $\tau\in [0, T]$.

Estimates \eq{rlpe1}, \eq{rlpe2}, \eq{rlpe3}, and \eq{estr2der}-\eq{estr2}
allows one to apply the standard technique of construction of the inverse to
the operator of a boundary problem for PDO to problem \eq{afk3}-\eq{abc3}.
This technique is based on the construction of an appropriate
partition of unity, localization and patching of an approximate inverse
from local inverses; for the realization of this general scheme for many
classes of PDO see Levendorski\v{i} (1993). In op. cit., boundary value problems
in $L_p$-based spaces were considered whereas here we need corresponding results
for $C^s$-based spaces. This modification is straightforward: see e.g. the modification
in Barndorff-Nielsen and
Levendorski\v{i} (2001), for a different class of PDO.

In the result, we obtain that $v$, the continuous solution to problem
\eq{afk3}-\eq{abc3}, which admits estimate \eq{boundf}, exists and
it is unique.
Moreover, it is of the class $C^{2, 1}$ in the half-plane $\tau>0$, and
satisfies estimate
\begin{equation}\label{estv}
\sup_{\rn\times [0, T]}|e^{-\om |y|}v(y, \tau)|\le C \sup_{\rn}
|e^{-\om |y|}g(y)|,
\end{equation}
where $C$ depends on $T$, $\ka$, $\tthet$ and the constants in estimates for
$\psi$ and $r$.
 By making the inverse changes of variables
and unknowns, we obtain (i) and (ii).

If $\ka$ is not diagonalizable, then prior to the change of
variables $x=Cy$, an additional change of variables $x_j\mapsto e^{\rho_j}x_j, j=1,\ldots, n$,
where $\rho_j>0$, is needed. The $\ka$ will be replaced by $\ka-{\rm diag} (\rho_j)$,
which generically has pairwise distinct eigenvalues and hence, diagonalizable; $\tthet$ will change
as well, and $\psi(D_x)$ becomes $\psi(e^{-\rho_1\tau} D_{x_1},\ldots, e^{-\rho_n\tau} D_{x_n}).$
After that we make the same changes of variables (using the new $\ka$ and $\tthet$).

To prove (iii), we take $\om_1\in (\om, \la)$, and apply the argument above
starting with $g-g^m$ instead of $g$ and $\om_1$ instead of $\om$. Since
\[
\sup_{\rn} |e^{-\om_1|x|}(g(x)-g^m(x))|\to 0\quad {\rm as}\ m\to \infty,
\]
estimate \eq{estv} implies that
\[
\sup_{\rn\times [0, T]}|e^{-\om_1|x|}(f(x, t)-f^m(x, t))|\to 0\quad {\rm as}\ m\to \infty,
\]
which proves (iii).

It remains to prove (iv). We have seen that for a continuous $g$
with compact support, $f$ is continuous in the half-plane $t<T$,
and of the class $C^{2, 1}$ in the open half-plane. Moreover,
$f(x, \tau)$ decays faster than $e^{-\om |x|}$ as $x\to \infty$,
for any $\om>-\la$ (notice that a continuous $g$ of the compact
support satisfies \eq{boundg} for any $\om$, and in order that the
proof of estimate \eq{boundf} remain valid, we may use any
$\om>-\la$, negative ones in particular). Further, $r$ is
continuous and semi-bounded from below. These conditions are more
than sufficient for the Feynman-Kac theorem to be applicable (for
instance, at this stage, we can repeat the proof on p.274 in
Rogers and Williams (1994)), which gives (iv). \theor{thmfk1} has
been proved.

\vskip0.2cm
\centerline{REFERENCES} \vskip0.1cm \noindent
{\sc Ahn, D.-H., R.F.~Dittmar, and A.R.~Gallant} (2002): Quadratic
term structure

models: theory and evidence, {\em Review of Financial Studies}
15:1, 243-288.

\vskip0.2cm \noindent
{\sc Ahn, D.-H., R.F.~Dittmar, A.R.~Gallant, and Bin Gao} (2003):
Pure bred or

hybrid?: Reproducing the volatility in term structure dynamics,
{\em Journal of Econo-

metrics} 116 (2003), 147--180.

\vskip0.1cm
\noindent
{\sc Barndorff-Nielsen, O.E.} (1998): Processes of Normal Inverse
Gaussian Type,

{\em Finance and Stochastics} 2, 41--68.

\vskip0.1cm \noindent {\sc Barndorff-Nielsen, O.E. and W.~ Jiang}
(1998): An initial analysis of some Ger-

man stock price series, Working Paper Series 15. Aarhus: CAF Univ.
of Aarhus/Aarhus

School of Business.

\vskip0.1cm \noindent {\sc Barndorff-Nielsen, O.E.,  and
S.~Levendorski\v{i}} (2001): Feller Processes of Nor-

mal Inverse Gaussian Type, {\em Quantitative Finance} 1, 318-331.

\vskip0.1cm \noindent {\sc Barndorff-Nielsen, O.E., E.~Nicolato,
and N.~Shephard} (2002): Some recent

developments in stochastic volatility modelling, {\em Quantitative
Finance} 2, 11-23.

\vskip0.1cm \noindent {\sc Barndorff-Nielsen, O.E.  and
N.~Shephard} (2001a): Normal modified stable

processes, Working paper MaPhySto: Aarhus University

\vskip0.1cm \noindent {\sc Barndorff-Nielsen, O.E. and
N.~Shephard} (2001b): Non-Gaussian Ornstein-

Uhlenbeck- based models and some of their uses in financial
economics (with discus-

sion), {\em J. Royal. Stat. Soc.} 63, 167-241.

\vskip0.1cm \noindent {\sc Bouchaud, J-P. and M.~Potters, M.}
(1997): {\em Theory of Financial Risk.}  Al\'ea-

 Saclay, Eurolles: Paris



\vskip0.1cm \noindent {\sc Boyarchenko, S.I. and
S.Z.~Levendorski\v{i}} (2000): Option pricing for truncated

L\'evy processes, {\em Intern. Journ.  Theor. and Appl. Finance}
3, 549-552.

\vskip0.1cm \noindent {\sc Boyarchenko, S.I. and
S.Z.~Levendorski\v{i}} (2002a): Perpetual American options

under L\'evy processes, {\em SIAM J. Control and Optimization} 40,
1663-1696.

\vskip0.1cm \noindent {\sc Boyarchenko, S.I. and
S.Z.~Levendorski\v{i}} (2002b): {\em Non-Gaussian
Merton-Black-

Scholes theory}. World Scientific: Singapore

\vskip0.1cm \noindent {\sc Boyarchenko, S.I. and
S.Z.~Levendorski\v{i}} (2002c): Barrier options and touch and

out options under regular L\'evy processes of exponential type,
{\em Annals of Applied Pro-

bability}, 12:4, 1261--1298.

\vskip0.1cm \noindent {\sc Boyarchenko, S.I. and
S.Z.~Levendorski\v{i}} (2002d): Pricing of perpetual Bermudan

options, {\em Quantitative Finance}, 2:6, 432--442.

\vskip0.1cm \noindent {\sc Carr,  P.,  H.~ Geman, D.B.~ Madan, M.~
Yor} (2002): The fine structure of asset

 returns: an empirical
investigation, {\em Journal of Business} 75, 305-332.

\vskip0.1cm \noindent {\sc Chacko, G., and S.~Das} (2002): Pricing
interest rate derivatives: a general approach,

{\em Review of Financial Studies} 15:1, 195-241.

\vskip0.1cm \noindent {\sc Chen, L. and H.V.~Poor} (2002): A
general characterization of quadratic term struc-

ture models, Working paper

\vskip0.1cm \noindent {\sc Chen, P. and O.~Scaillet} (2002):
Linear-quadratic jump-diffusion modeling with

application to stochastic volatility, Working paper

\vskip0.1cm \noindent {\sc Collin-Dufresne, P., R.S.~Goldstein,
and C.~Johnes} (2003): Identification

and estimation of ``maximal" affine term structure models: an
application to stochastic

volatility, Working paper

\vskip0.1cm \noindent {\sc Cont, R., M.~Potters,  and
J.-P.~Bouchaud} (1997): Scaling in stock market data:

stable laws and beyond, in  {\em Scale Invariance and beyond
(Proceedings of the CNRS

Workshop on Scale Invariance, Les Houches, March 1997)}. B.
Dubrulle, F. Graner,

and D. Sornette, eds., Springer: Berlin, 75--85.



\vskip0.1cm \noindent {\sc Das, S.R.} (2002): The surprise
element: jumps in interest rates, {\em Journal of Economet-

rics} 106, 27-65.

\vskip0.1cm \noindent {\sc Duffie, D., and R.~Kan} (1996): A
yield-factor model of interest rates, {\em Mathematical

Finance} 6, 376-406.

\vskip0.1cm \noindent {\sc Duffie, D., J.~Pan, and K.~Singleton}
(2000): Transform analysis and option

pricing for affine jump-diffusions, {\em Econometrica} 68,
1343-1376.

\vskip0.1cm \noindent {\sc Duffie, D., D.~ Filipovi\'c, and
W.~Schachermayer} (2002): Affine processes and

applications in Finance, {\em Annals of Applied Probability}



\vskip0.1cm \noindent {\sc Eberlein, E., U.~ Keller, and
K.~Prause} (1998): New insights into smile, mispric-

ing and value at risk: The hyperbolic model, {\em Journ. of
Business} 71, 371--406.

\vskip0.1cm \noindent {\sc Eberlein, E., and S.~Raible} (1999):
Term structure models driven by general L\'evy

process, {\em Mathematical Finance} 9 , 31-53.

\vskip0.1cm \noindent {\sc Eberlein, E., and F.~Ozkan}  (2001):
The defaultable L\'evy Term Structure: Ratings

and Restructuring, Preprint 71, University of Freiburg.

\vskip0.1cm \noindent {\sc Fama, E.F.} (1965): The behavior of
stock market prices, {\em Journ. of Business} 38, 34--105.



\vskip0.1cm \noindent {\sc Don H. Kim} (2003): Time-varying risk
and return in the quadratic-gaussian model of

the term structure, Working paper


\vskip0.1cm \noindent {\sc Koponen, I.} (1995): Analytic approach
to the problem of convergence of truncated

L\'evy flights towards the Gaussian stochastic process, {\em
Physics Review E} 52, 1197--1199.

\vskip0.1cm \noindent {\sc Kudryavzev, O.E., and
S.Z.~Levendorski\v{i}} (2002): Comparative study of first

touch digitals: normal inverse gaussian vs. gaussian modelling,
Working paper

MaPhySto: Aarhus, October 2002.

\vskip0.1cm \noindent {\sc Levendorski\v{i}, S.} (1993):
 {\em Degenerate elliptic equations}.
Mathematics and its Applica-

tions, 258. Kluwer Academic Publishers Group: Dordrecht

\vskip0.1cm \noindent {\sc Levendorski\v{i}, S.} (2003): The
American put and European options near expiry, under

L\'evy processes, Working paper

\vskip0.1cm \noindent {\sc Madan, D.B.,  P.~Carr,  and E.C.~Chang}
(1998): The variance Gamma process and

option pricing, {\em European Finance Review} 2, 79--105.





\vskip0.1cm \noindent {\sc Mandelbrot, B.B.} (1963): The variation
of certain speculative prices, {\it Journ. of

Business}  36, 394-419.

\vskip0.1cm \noindent {\sc Matacz, A.} (2000): Financial modelling
and option theory with the truncated L\'evy

process, {\em Intern. Journ.  Theor. and Appl. Finance} 3:1,
143-160.



\vskip0.1cm \noindent {\sc Rogers, L.C.G., and D.~Williams}
(1994): {\em Diffusions, Markov processes, and mar-

tingales"}, John Wiley and Sons: West Sussex UK

\vskip0.1cm \noindent {\sc K. Sato} (1999), {\em L\'evy processes
and infinitely divisible distributions}, Cambridge Univer-

sity Press: Cambridge

\vskip0.1cm \noindent {\sc Singleton, K.} (1999): Estimation of
affine asset pricing models unsing the empirical

characteristic function, {\em Journal of Econometrics} 102,
111--141

\newpage
\begin{figure}
 \scalebox{0.8}
{\includegraphics{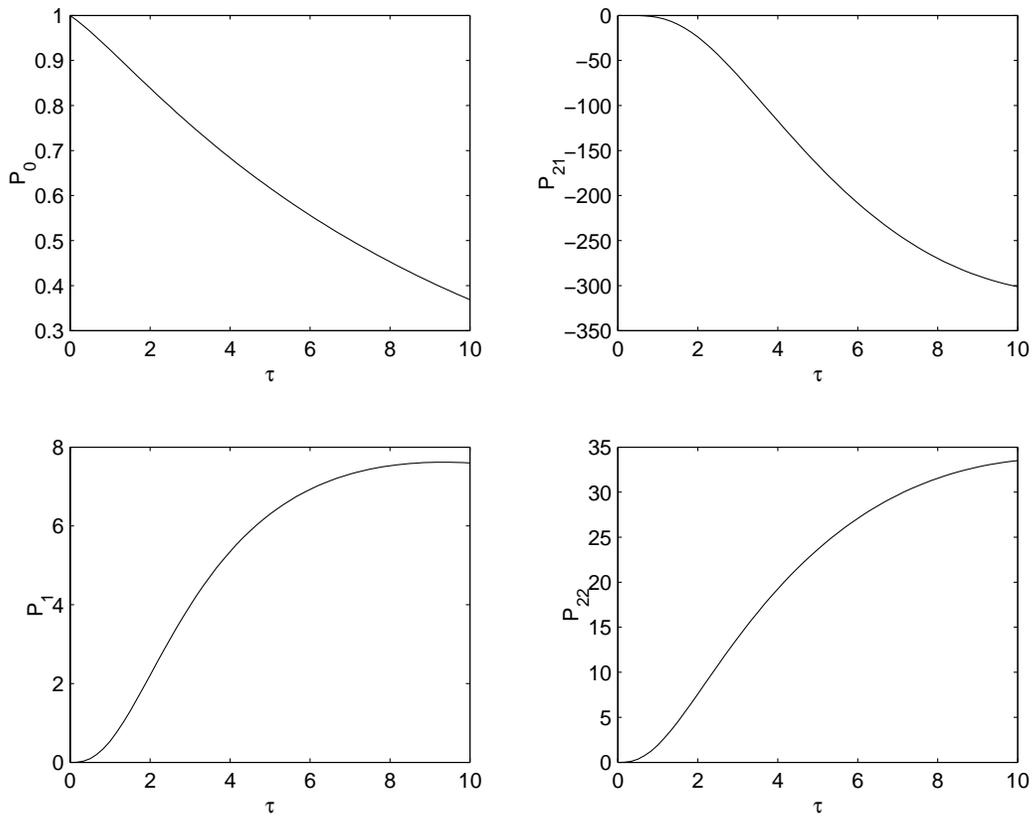}}
 \caption{Components of the asymptotics of
  the bond price. Parameters:
  $r=x^2=0.0625, \tthet=0.06, \ka=0.3, \mu=0, \sg^2=0.08$.}
 \end{figure}

 \newpage
\begin{figure}

 \scalebox{0.8}
{\includegraphics{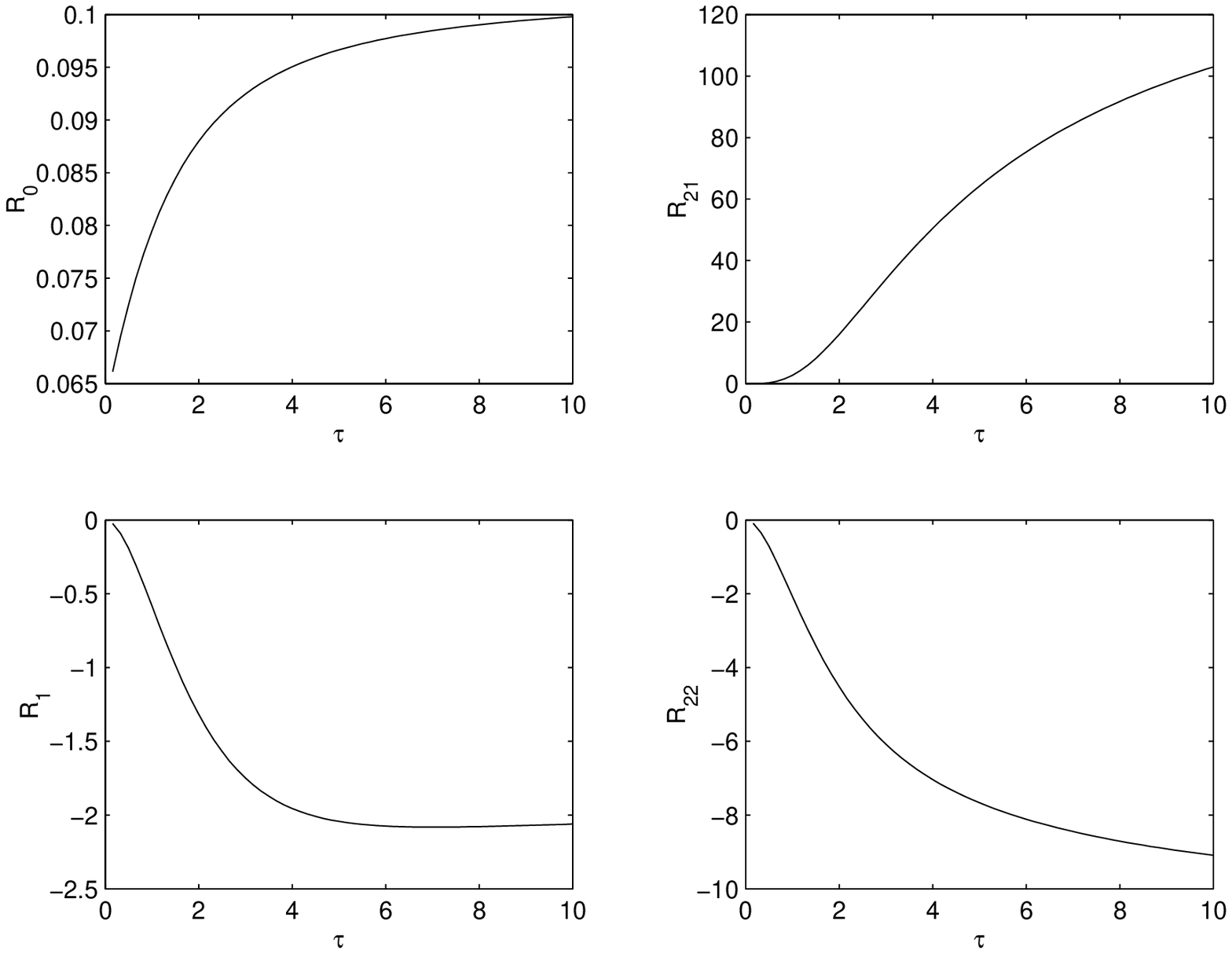}}
 \caption{Components of the asymptotics of  the yield.
  Parameters:
   $r=x^2=0.0625, \tthet=0.06, \ka=0.3, \mu=0, \sg^2=0.08$.}
 \end{figure}
 \newpage
 \begin{figure}
 \scalebox{0.8}
{\includegraphics{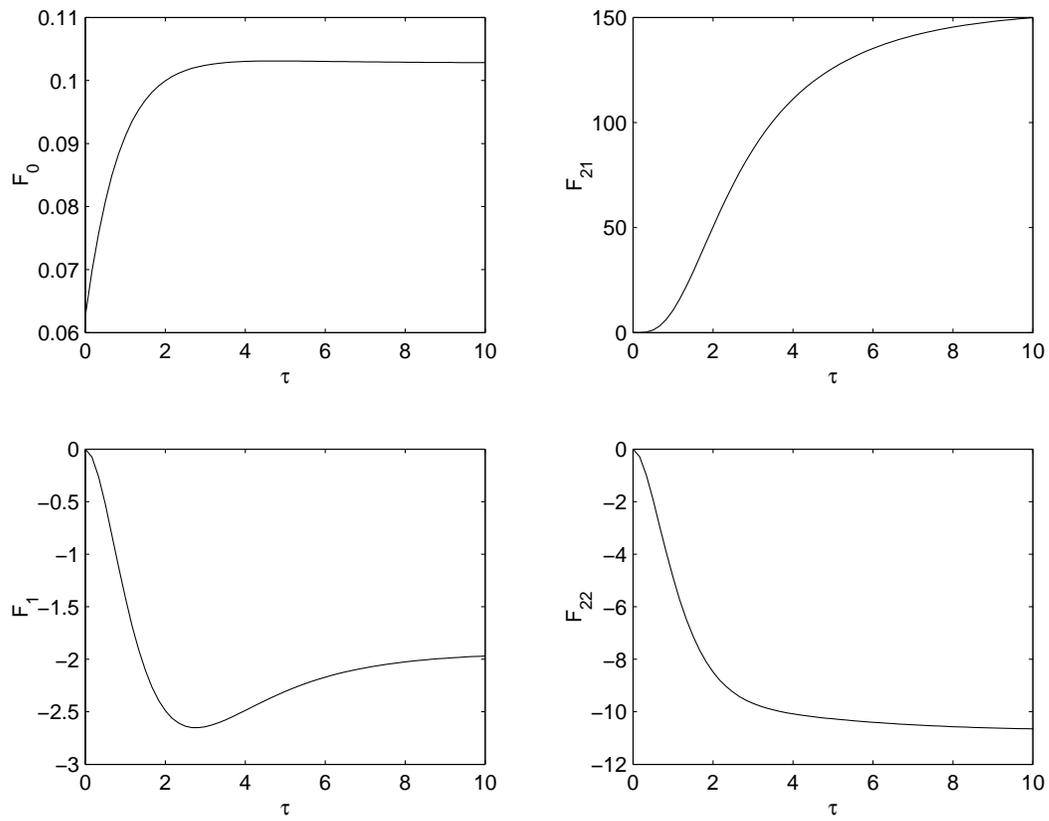}}
 \caption{Components of the asymptotics of forward rate.
  Parameters:
   $r=x^2=0.0625, \tthet=0.06, \ka=0.3, \mu=0, \sg^2=0.08$.}
 \end{figure}

 \newpage
\begin{figure}
 \scalebox{0.8}
{\includegraphics{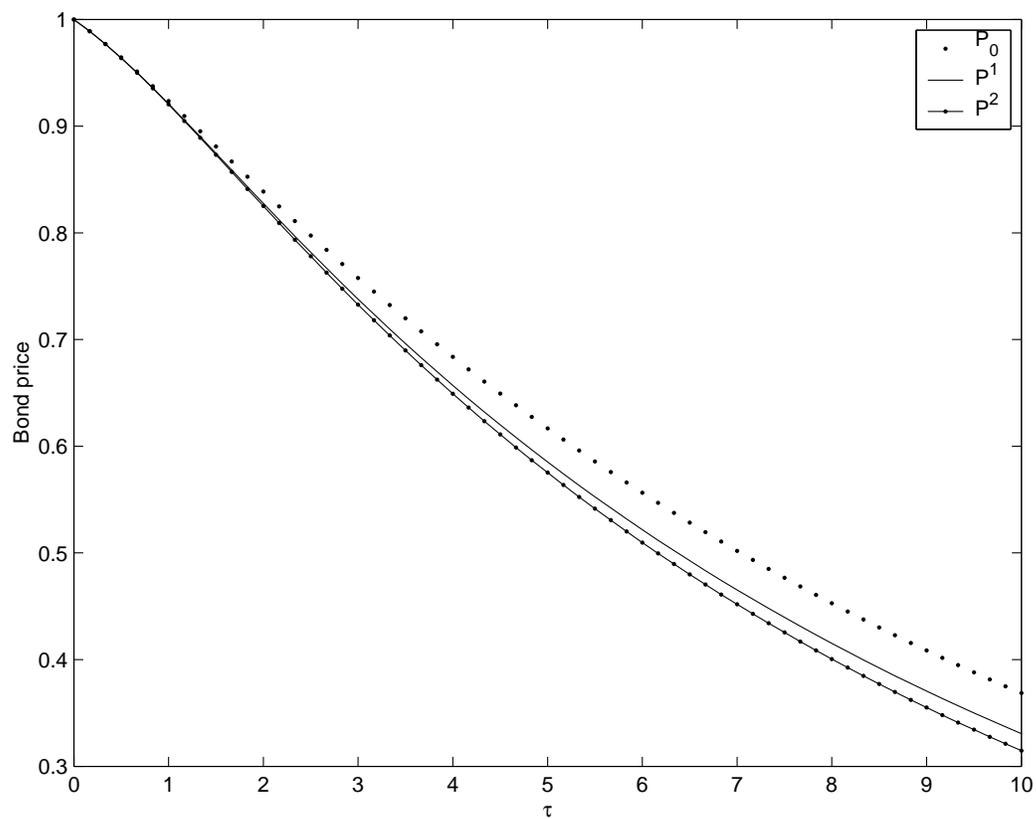}}
 \caption{Bond price: leading term, first and second
 approximations.
  Parameters:
  $r=x^2=0.0625, \tthet=0.06, \ka=0.3, \mu=0, \sg^2=0.08,
K_3=-0.005, K_4=2.5\cdot 10^{-4}$.}
 \end{figure}

\newpage
\begin{figure}
 \scalebox{0.8}
{\includegraphics{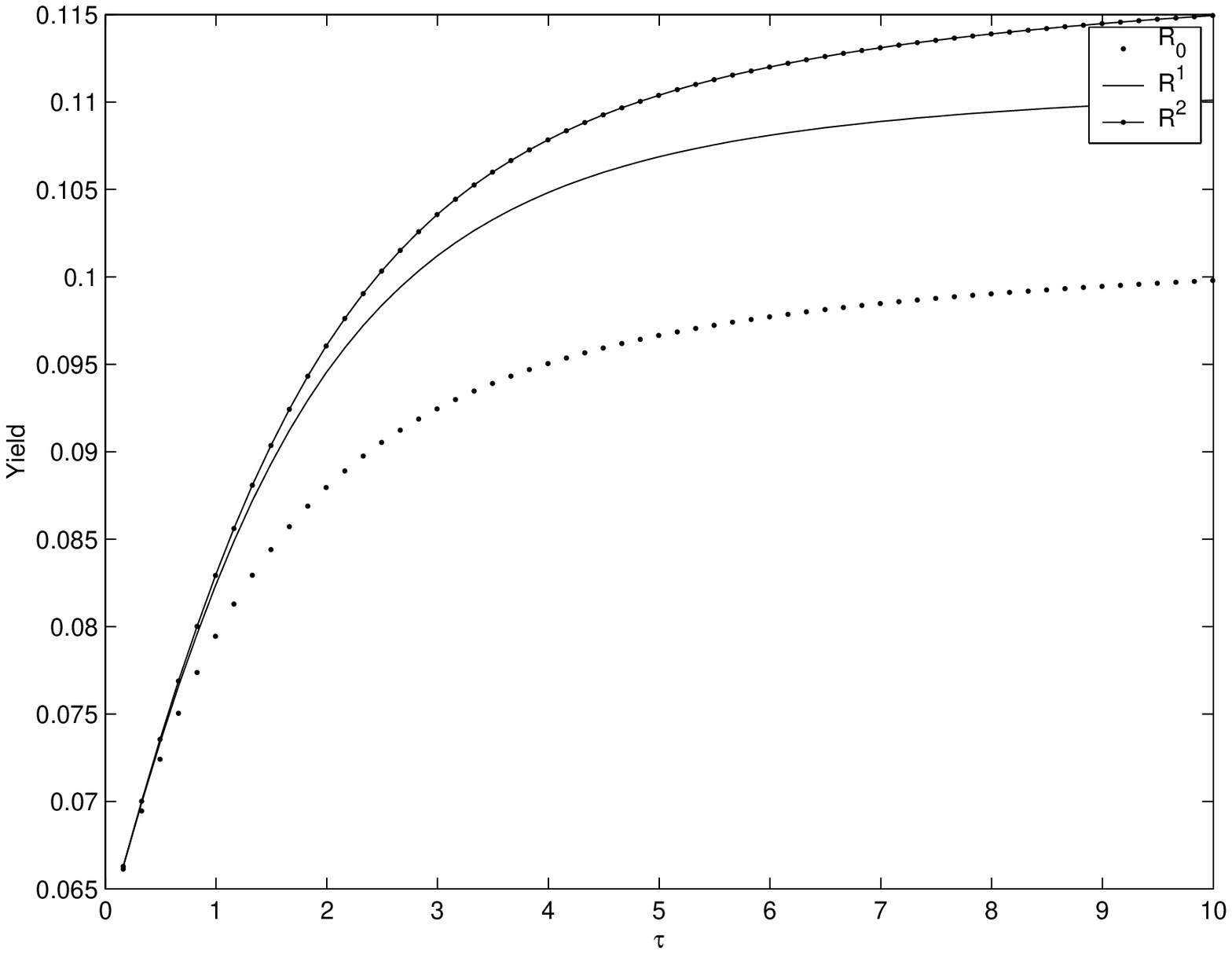}}
 \caption{Yield: leading term, first and second
 approximations.
 Parameters:
  $r=x^2=0.0625, \tthet=0.06, \ka=0.3, \mu=0, \sg^2=0.08,
K_3=-0.005, K_4=2.5\cdot 10^{-4}$.}
 \end{figure}
 \newpage
 \begin{figure}

 \scalebox{0.8}
{\includegraphics{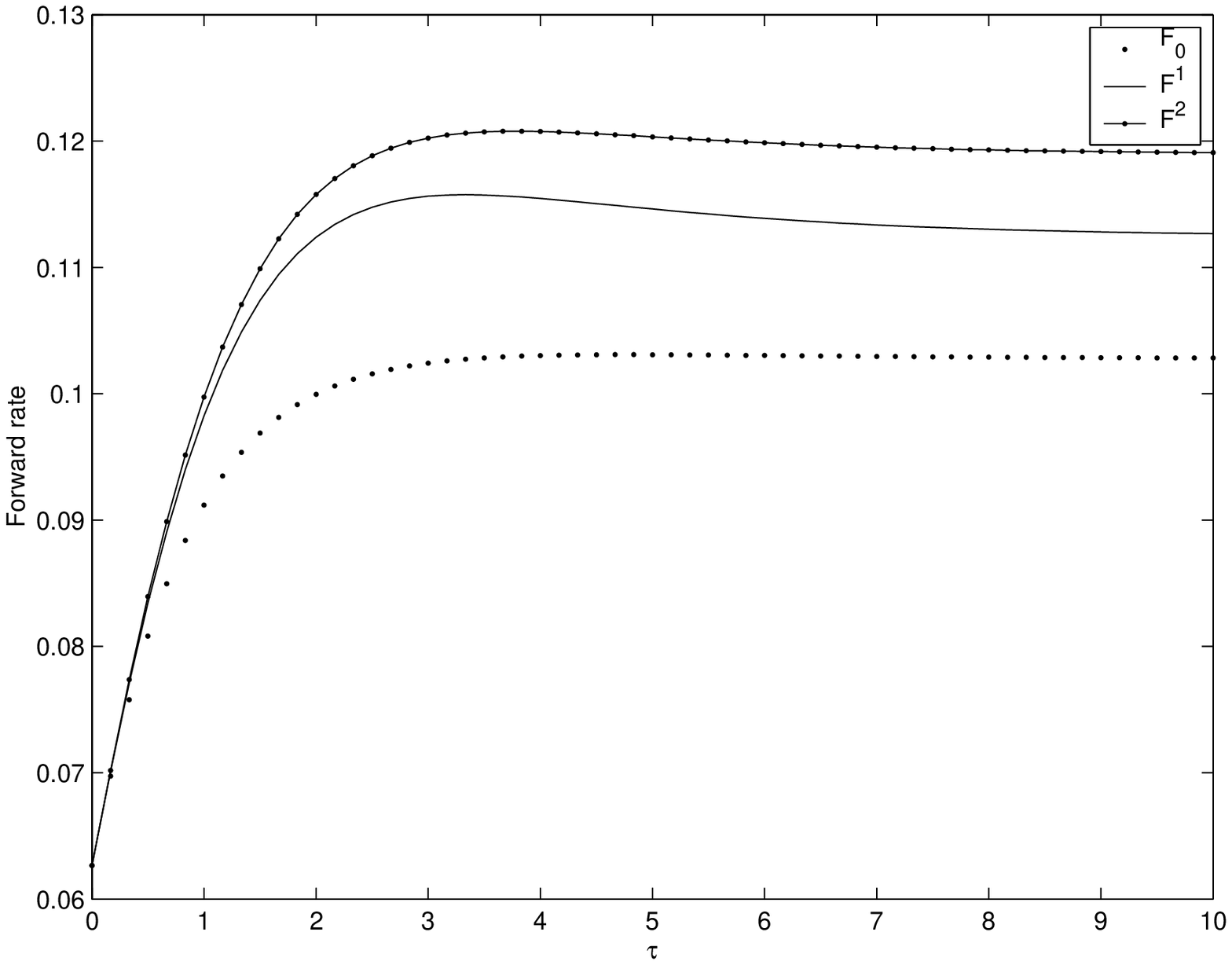}}
 \caption{Forward rate: leading term,
 first and second approximations.
 Parameters:
  $r=x^2=0.0625, \tthet=0.06, \ka=0.3, \mu=0, \sg^2=0.08,
K_3=-0.005, K_4=2.5\cdot 10^{-4}$.}
 \end{figure}
\newpage
 \begin{figure}
 \scalebox{0.8}
{\includegraphics{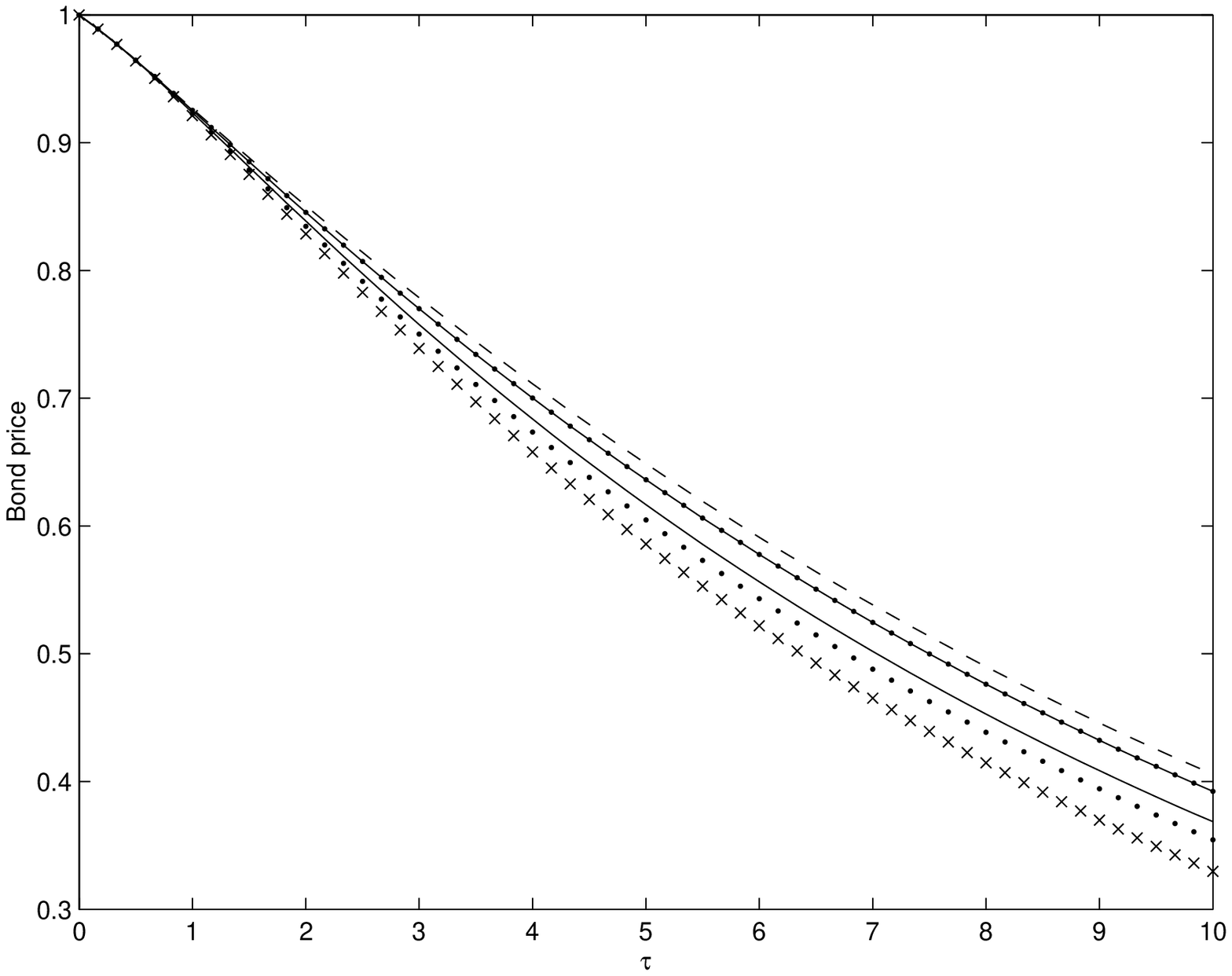}}
 \caption{Bond price: dependence on skewness. Parameters:
  $r=x^2=0.0625, \tthet=0.06, \ka=0.3, \mu=0, \sg^2=0.08,
 K_4=2\cdot 10^{-4}$. Dashes: $K_3=5\cdot 10^{-3}$; dotted
line: $K_3=2.5\cdot 10^{-3}$; dots: $K_3=-2.5\cdot 10^{-3}$;
crosses: $K_3=-5\cdot 10^{-3}$. The solid line is the Gaussian
curve.}
\end{figure}
\vfill
 \newpage
\begin{figure}
 \scalebox{0.8}
{\includegraphics{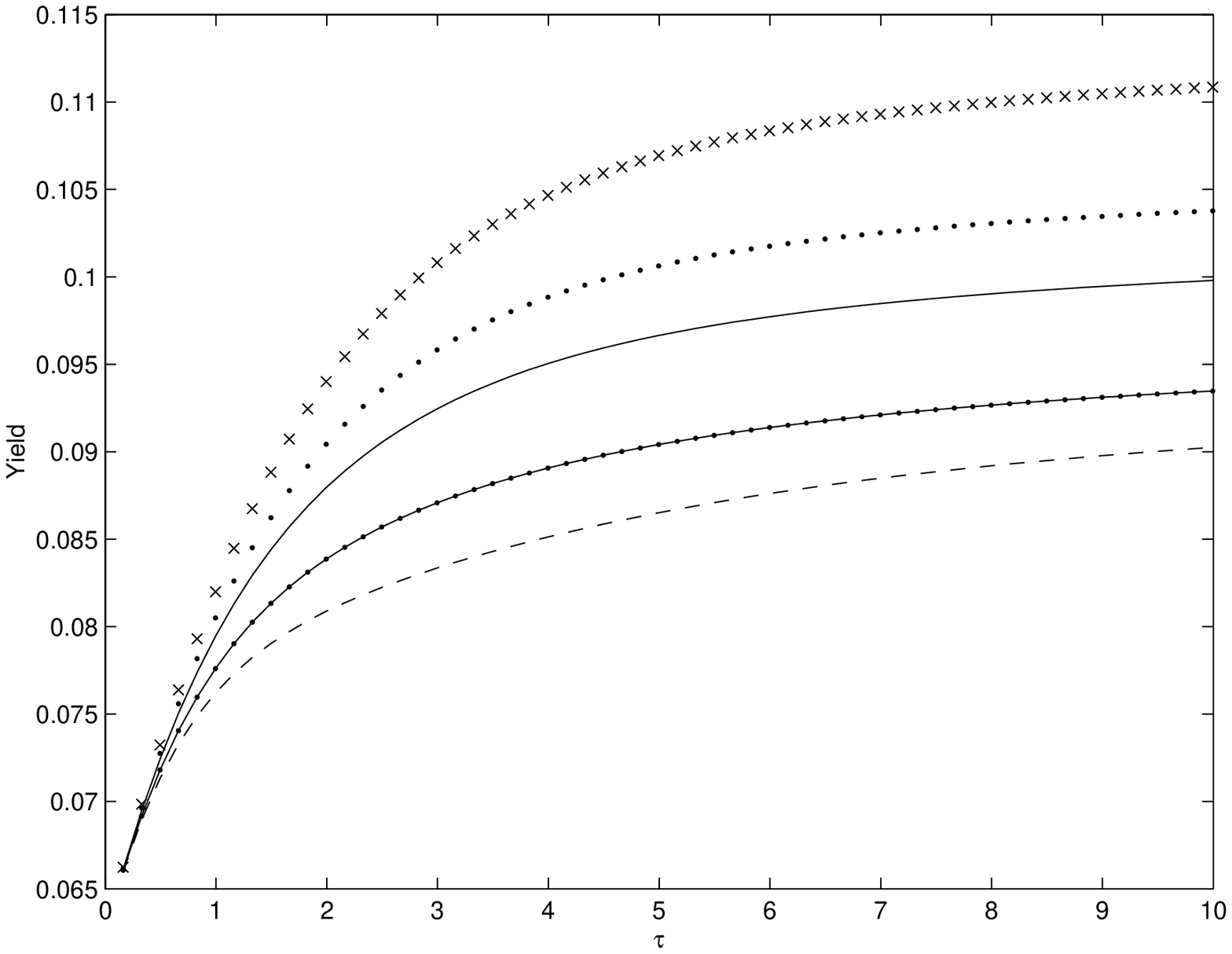}}
 \caption{Yield: dependence on skewness. Parameters:
  $r=x^2=0.0625, \tthet=0.06, \ka=0.3, \mu=0, \sg^2=0.08,
 K_4=2\cdot 10^{-4}$. Dashes: $K_3=5\cdot 10^{-3}$; dotted
line: $K_3=2.5\cdot 10^{-3}$; dots: $K_3=-2.5\cdot 10^{-3}$;
crosses: $K_3=-5\cdot 10^{-3}$. The solid line is the Gaussian
curve.}
 \end{figure}
 \vfill
 \newpage
\begin{figure}
 \scalebox{0.8}
{\includegraphics{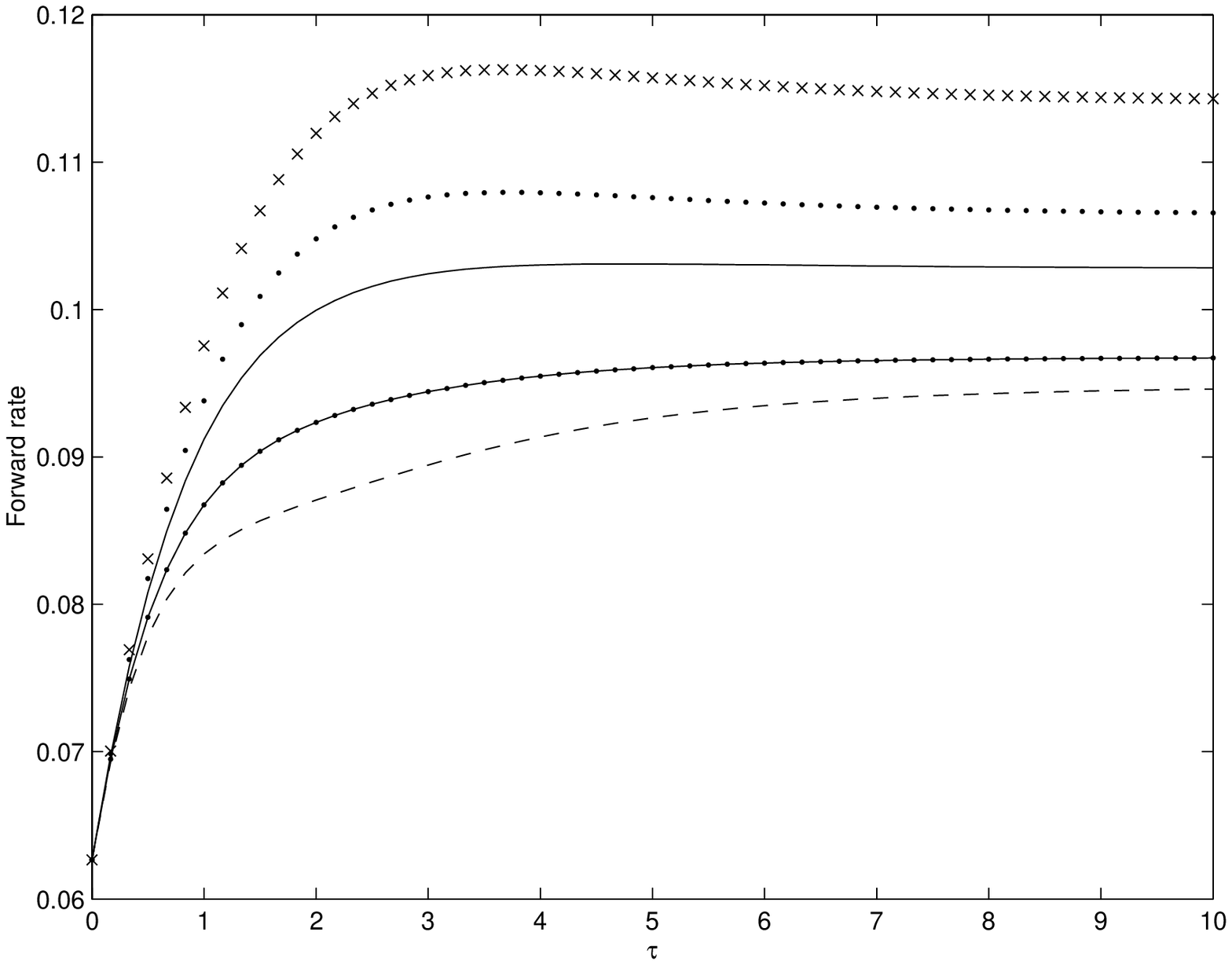}}
 \caption{Forward rate: dependence on skewness. Parameters:
  $r=x^2=0.0625, \tthet=0.06, \ka=0.3, \mu=0, \sg^2=0.08,
 K_4=2\cdot 10^{-4}$. Dashes: $K_3=5\cdot 10^{-3}$; dotted
line: $K_3=2.5\cdot 10^{-3}$; dots: $K_3=-2.5\cdot 10^{-3}$;
crosses: $K_3=-5\cdot 10^{-3}$. The solid line is the Gaussian
curve.}
 \end{figure}

\end{document}